\documentclass[final,3p,times,twocolumn]{elsarticle}
\usepackage{amssymb}
\usepackage{hyperref}
\usepackage{subcaption}
\usepackage{lineno}
\usepackage{amsmath}
\usepackage{cleveref}
\Crefname{figure}{Fig.}{Figs.}
\graphicspath{{Figures/}}
\journal{NIM A}
\begin{document}
\begin{frontmatter}
\title{Study of time and energy resolution of an ultra-compact sampling\\ calorimeter (RADiCAL) module at EM shower maximum\\ over the energy range 25~GeV~$\leq$ E $\leq$~150~GeV}
\author[label4]{Carlos Perez-Lara}
\author[label3,label5]{James Wetzel}
\author[label3,label5]{Ugur Akgun}
\author[label10]{Thomas Anderson}
\author[label9]{Thomas Barbera}
\author[label5]{Dylan Blend}
\author[label8]{Kerem Cankocak}
\author[label1,label11]{Salim Cerci}
\author[label10]{Nehal Chigurupati}
\author[label10]{Bradley Cox}
\author[label5]{Paul Debbins}
\author[label10]{Max Dubnowski}
\author[label6]{Buse Duran}
\author[label8]{Gizem Gul Dincer}
\author[label6]{Selbi Hatipoglu}
\author[label7]{Ilknur Hos}
\author[label11]{Bora Isildak}
\author[label9]{Colin Jessop}
\author[label4]{Ohannes Kamer Koseyan}
\author[label11]{Ayben Karasu Uysal}
\author[label11]{Reyhan Kurt}
\author[label6]{Berkan Kaynak}
\author[label10]{Alexander Ledovskoy}
\author[label5]{Alexi Mestvirishvili}
\author[label5]{Yasar Onel}
\author[label6]{Suat Ozkorucuklu}
\author[label5]{Aldo Penzo}
\author[label6]{Onur Potok}
\author[label9]{Daniel Ruggiero}
\author[label9]{Randal Ruchti}
\author[label1,label11]{Deniz Sunar Cerci}
\author[label6]{Ali Tosun}
\author[label9]{Mark Vigneault}
\author[label9]{Yuyi Wan}
\author[label9]{Mitchell Wayne}
\author[label11]{Taylan Yetkin}
\author[label2]{Liyuan Zhang}
\author[label2]{Renyuan Zhu}
\author[label6]{Caglar Zorbilmez}

\affiliation[label1]{organization={Adiyaman University},
             city={Adiyaman},
             country={Turkiye}}
\affiliation[label2]{organization={California Institute of Technology},
             city={Pasadena},
             country={CA USA}}
\affiliation[label3]{organization={Coe College},
             city={Cedar Rapids},
             state={IA},
             country={USA}}
\affiliation[label4]{organization={Hofstra University},
             city={Hempstead},
             state={NY},
             country={USA}}
\affiliation[label5]{organization={University of Iowa},
             city={Iowa City},
             state={IA},
             country={USA}}
\affiliation[label6]{organization={Istanbul University},
             city={Istanbul},
             country={Turkiye}}
\affiliation[label7]{organization={Istanbul University - Cerrahpasa},
             city={Istanbul},
             country={Turkiye}}
\affiliation[label8]{organization={Istanbul Technical University},
             city={Istanbul},
             country={Turkiye}}
\affiliation[label9]{organization={University of Notre Dame},
             city={Notre Dame},
             state={IN},
             country={USA}}
\affiliation[label10]{organization={University of Virginia},
             city={Charlottesville},
             state={VA},
             country={USA}}
\affiliation[label11]{organization={Yildiz Technical University},
             city={Istanbul},
             country={Turkiye}}

\begin{abstract}
The RADiCAL Collaboration is conducting R\&D on high performance electromagnetic (EM) calorimetry to address the challenges expected in future collider experiments under conditions of high luminosity and/or high irradiation (FCC-ee, FCC-hh and fixed target and forward physics environments). Under development is a sampling calorimeter approach, known as RADiCAL modules, based on scintillation and wavelength-shifting (WLS) technologies and photosensor, including SiPM and SiPM-like technology. The modules discussed herein consist of alternating layers of very dense (W) absorber and scintillating crystal (LYSO:Ce) plates, assembled to a depth of 25 $X_0$. The scintillation signals produced by the EM showers in the region of EM shower maximum (shower max) are transmitted to SiPM located at the upstream and downstream ends of the modules via quartz capillaries which penetrate the full length of the module. The capillaries contain DSB1 organic plastic WLS filaments positioned within the region of shower max, where the shower energy deposition is greatest, and fused with quartz rod elsewhere. The wavelength shifted light from this spatially-localized shower max region is then propagated to the photosensors. This paper presents the results of an initial measurement of the time resolution of a RADiCAL module over the energy range 25 GeV $\leq$ E $\leq$ 150 GeV using the H2 electron beam at CERN. The data indicate an energy dependence of the time resolution that follows the functional form: $\sigma_{t} = a/\sqrt{E} \oplus b$, where a = 256 $\sqrt{GeV}$~ps and b = 17.5 ps.  The time resolution measured at the highest electron beam energy for which data was currently recorded (150 GeV) was found to be $\sigma_{t}$ = 27 ps.
\end{abstract}

%


\end{frontmatter}

\section{Introduction}
\label{Introduction}

    The R\&D objectives of the RADiCAL collaboration are focused on the development of precision EM calorimetry, based primarily on optical techniques, for future colliding beam experiments at facilities such as FCC-ee, FCC-hh \cite{fcc2019,fcc-ee2019,fcc-ee2019-2,fcc-hh2019,aleksa2019}, as well as fixed target and forward direction/small angle experiments. The effort is directed toward addressing the Priority Research Directions (PRD) for calorimetry listed in the US Department of Energy Basic Research Needs (BRN) workshop for HEP Instrumentation \cite{fleming2019}, as well as active participation in collaborative efforts currently in progress through ECFA DRD6 \cite{ecfa2019} and CPAD CRD9 programs \cite{cpad2019,rdc2019,rdc92019,wetzel2023}.

    Through the development of fast, radiation-hard scintillators, wavelength shifters, and optical transmission elements, performance goals include producing modular structures capable of achieving: an EM energy resolution of $\sigma_{E}/E = 10\%/\sqrt{E}~\oplus~0.3/E~\oplus~0.7\%$ \cite{aleksa2019}; a timing resolution of $\sigma_{t}~\leq~20$~ps at high energy \cite{ledovskoy2021}; and EM shower centroid position localization to within a few mm \cite{ledovskoy2021,ledovskoy2021-2}. Measurements of the radiation tolerance of various fluorescent and photosensing elements \cite{anderson2022} along with initial measurements of the energy resolution and timing resolution have been presented elsewhere \cite{anderson2022,anderson2022-2,wetzel2023-2} and remain under active development and further study. This paper presents first results from a beam test of the RADiCAL module concept for timing resolution over an extensive range of electron beam energies: $25~\text{GeV}~\leq~\text{E}~\leq~150~\text{GeV}$ \cite{perez2023,wetzel2023-3}.

\section{RADiCAL module under test}
\label{RMUT}

\begin{figure}
    	\begin{center}
    		\includegraphics[width=0.45\textwidth]{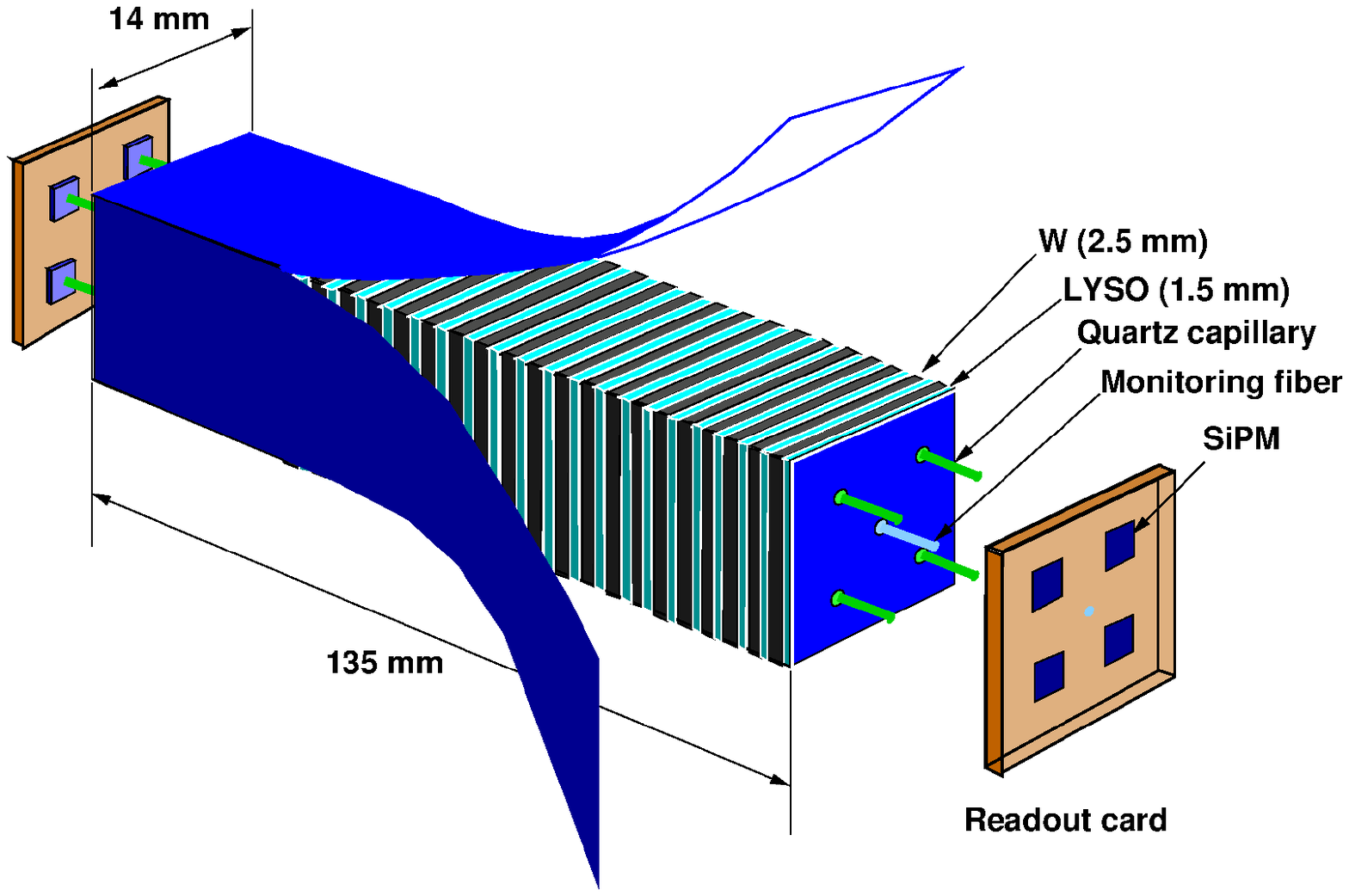}
    		\caption{A schematic of a RADiCAL module for ultra-compact EM calorimetry based upon interleaved layers of W and LYSO:Ce crystal in a Shashlik/Kebap-like structure and read out via specialized WLS quartz capillary elements which penetrate through the module to SiPM photosensors, positioned at both ends of the module. In this schematic, the beam enters the module from the upper left. The cross-sectional dimensions of the module are set by the Moli\`{e}re radius of the structure. }
    		\label{RadSchem}
    	\end{center}
    \end{figure}

\begin{figure}
    	\begin{center}
    		\includegraphics[width=0.45\textwidth]{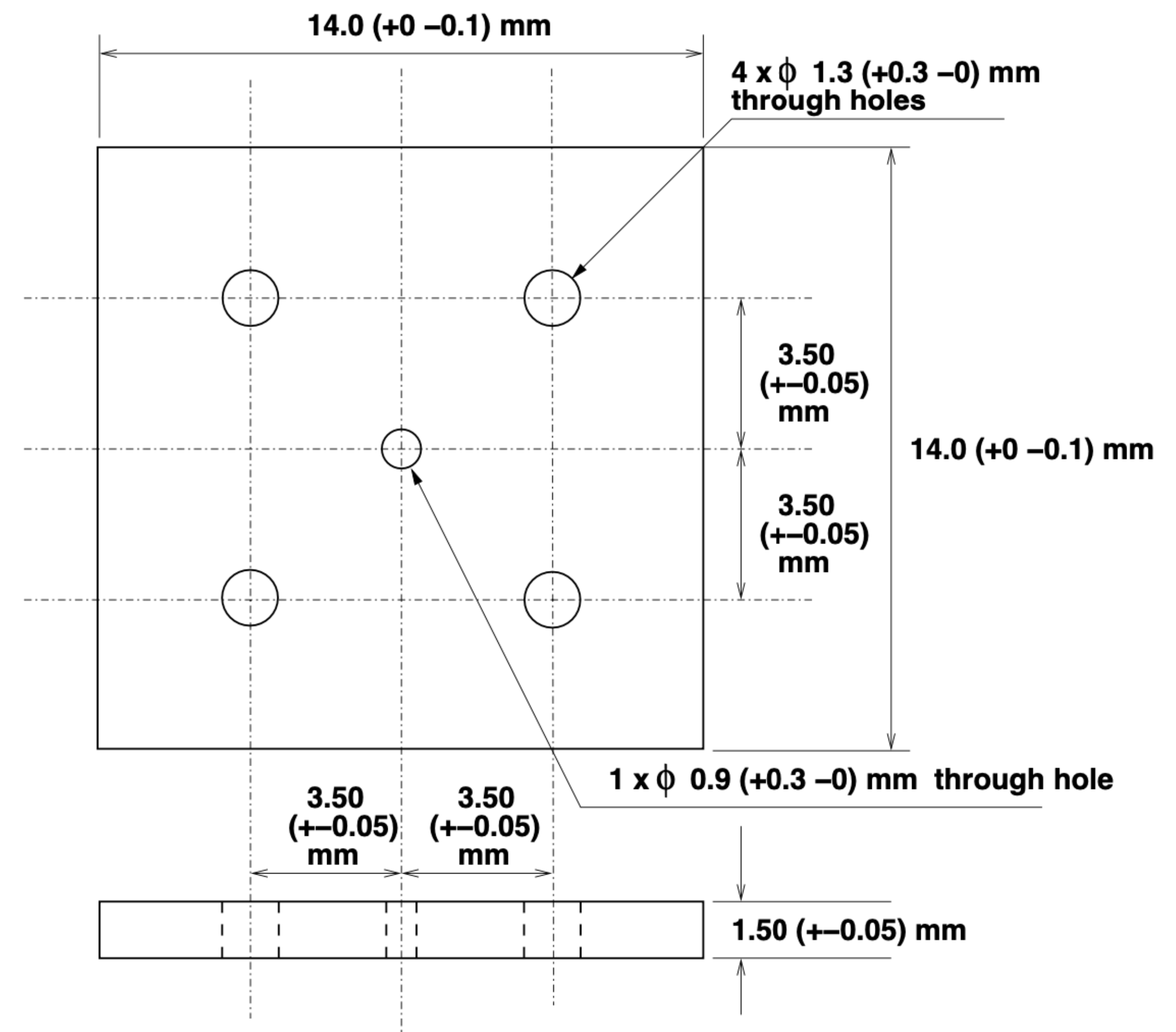}
    		\caption{Cross section of the RADiCAL module tiles, indicating the five locations (holes) for capillary waveguide insertion. Tungsten plates and LYSO:Ce scintillation plates had similar cross-sectional dimensions. The LYSO:Ce plates, as shown, had a thickness of 1.5 mm. The tungsten plates had a thickness of 2.5 mm. The central hole location, available for calibration or additional measurement, was not used in these tests reported here.}
    		\label{LysoPlate}
    	\end{center}
    \end{figure}

    For these studies, the RADiCAL prototype, Fig.~\ref{RadSchem}, is a Shashlik/Kebap-style module of overall dimensions 14 x 14 x 135 mm$^3$ that consists of a layered structure of 29 LYSO:Ce crystal plates of 1.5 mm thickness interleaved with 28 W plates of 2.5 mm thickness and assembled to a depth of $\approx 25~X_{0}$ and corresponding to an absorption length of $\approx 0.9~\lambda$. Thin Tyvek sheets were inserted between successive layers to act as reflective spacers to avoid optical absorption between LYSO:Ce and W surfaces. Overall, the structure has a radiation length $X_{0}$ = 5.4 mm and a Moli\'{e}re radius $R_{M}$~= 13.7 mm. Holes were drilled through five locations in all of the tiles as shown schematically in Fig.~\ref{LysoPlate}, to facilitate the penetration of quartz capillaries through the layered structure, each capillary containing wavelength shifter (WLS) filaments positioned strategically to facilitate energy or timing measurement.  SiPM photosensors were mounted on front-end electronic cards located at both upstream and downstream ends of the module to detect the optical signals from the capillaries.

\begin{figure}
    	\begin{center}
    		\includegraphics[width=0.45\textwidth]{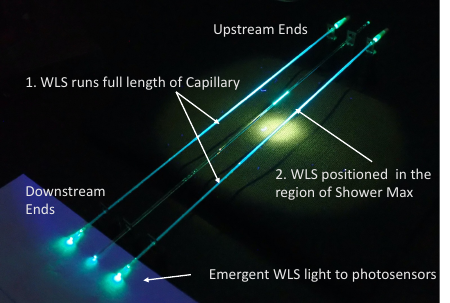}
    		\caption{Response of WLS Capillaries containg DSB1 waveshifter to UV light at 365 nm. Upper and lower are E-Type capillaries. Middle is a T-Type capillary.}
    		\label{Caps}
    	\end{center}
    \end{figure}

    To date, several types of WLS capillaries have been fabricated for the ongoing testing, as shown in Fig.~\ref{Caps}. They include: E-Type capillaries for energy measurement, where the capillary cores are filled either with a liquid wavelength shifter  or with a solid WLS filament that runs the full length of the module; and T-type capillaries for timing measurement, in which solid WLS filaments of short length are inserted into the capillary cores to a position corresponding to the region of EM shower max. The remaining core volume of such T-type capillaries are filled with quartz rods which are inserted and fused to the quartz walls of the capillaries, creating a solid quartz waveguides upstream and downstream of the WLS filaments.
    
\begin{figure}
    	\begin{center}
    		\includegraphics[width=0.45\textwidth]{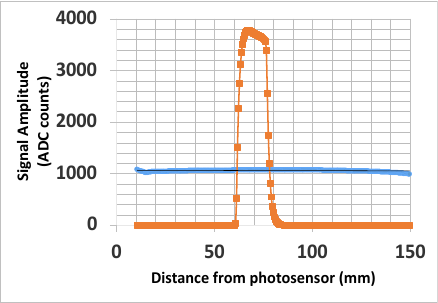}
    		\caption{Example of the optical response to LED light of 425 nm as a function of position for capillaries of E-type (blue) and T-type (orange) with the photodetector placed at the left (0 mm location).}
    		\label{Caps2}
    	\end{center}
    \end{figure}

    Using a custom-built capillary-tube tester, the optical responses of both E-Type and T-Type capillaries containing DSB1 wavelength shifters were assessed in laboratory bench tests \cite{DSB1}. DSB1 absorbs strongly at $\lambda = 425~nm$, emission peak at $\lambda = 495~nm$ and fast fluorescence decay $\tau = 3.5~ns$.  Fig.~\ref{Caps2} shows the response of such capillaries when the wavelength shifter is excited by an LED at $\lambda = 425~nm$. Fig.~\ref{Caps2} indicates that the response of an E-type capillary is very uniform over the full capillary length, desirable for longitudinal uniformity for energy measurement. Also shown is the highly localized response of the T-type, pronounced in the shower max region and negligible elsewhere, desirable for precision timing. It should be noted that the T-type capillaries also provide a localized measurement of the shower energy at shower max, important for shower position determination at that location.
    
\begin{figure}
    	\begin{center}
    		\includegraphics[width=0.45\textwidth]{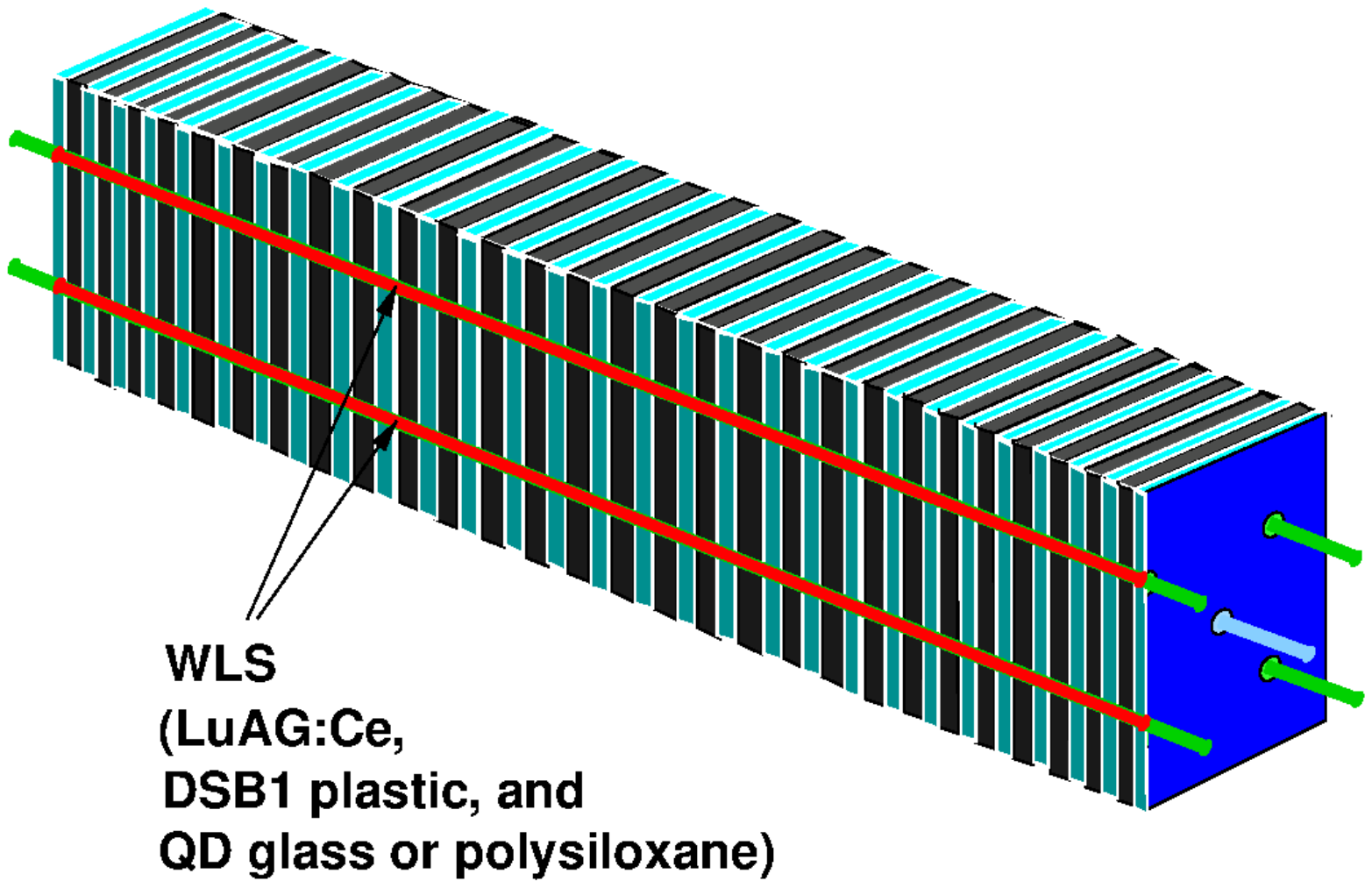}
    		\caption{RADiCAL module with E-type capillaries where the WLS runs the full length of the modules.}
    		\label{EType}
    	\end{center}
    \end{figure}

\begin{figure}
    	\begin{center}
    		\includegraphics[width=0.45\textwidth]{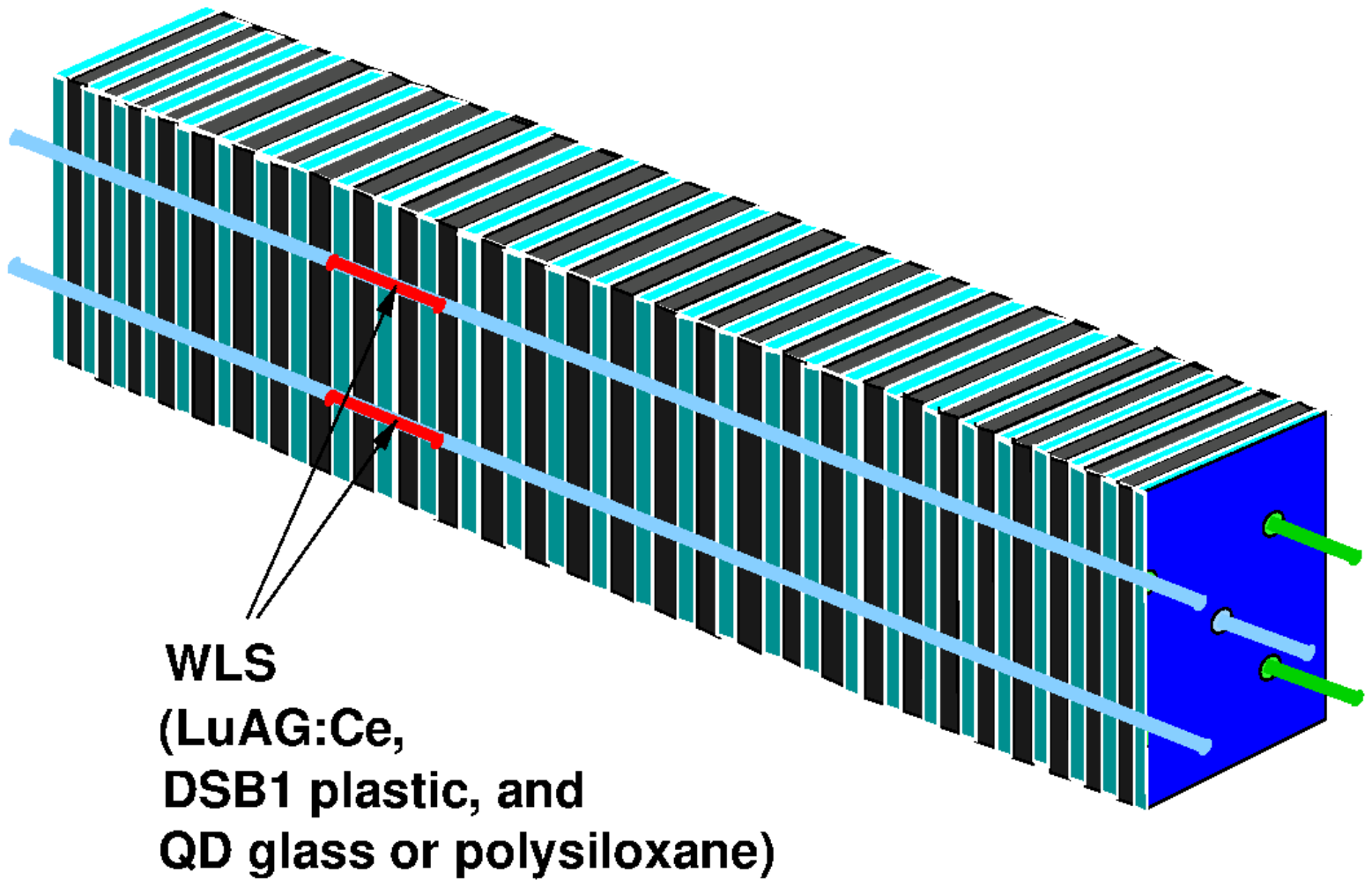}
    		\caption{RADiCAL module with T-type capillaries where the WLS is located in the region of shower max.}
    		\label{TType}
    	\end{center}
    \end{figure}

    Fig.~\ref{EType} and Fig.~\ref{TType} show schematics of a RADiCAL module instrumented with E-type and T-type capillaries appropriate for the testing of energy or timing resolution. Ultimately, an experimental application would utilize both capillary types positioned strategically within the module. This evolution is discussed in Section~\ref{section:ORD}.
    
\begin{figure}
    	\begin{center}
            \includegraphics[width=0.45\textwidth]{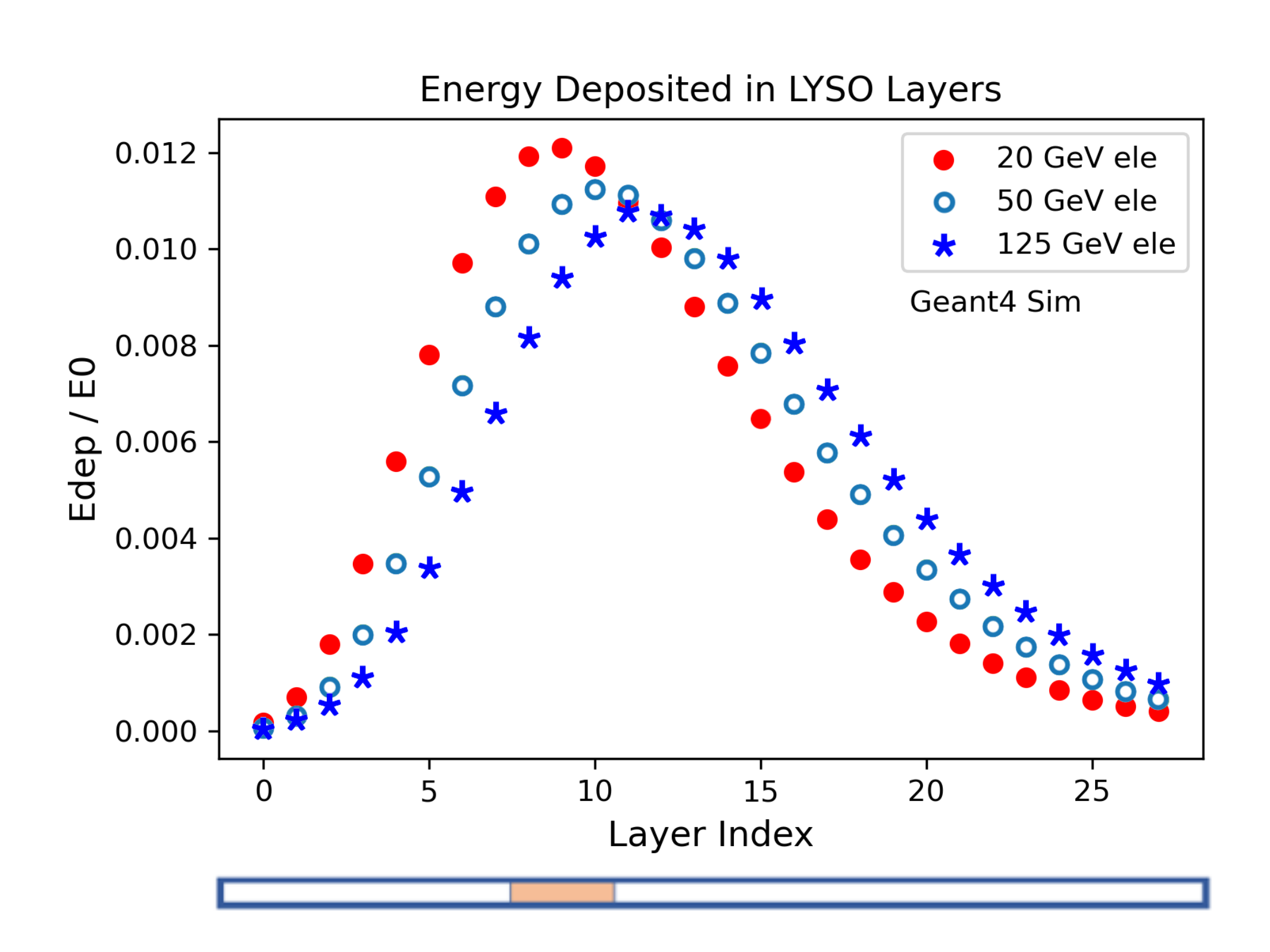}
    		\caption{GEANT4 simulation of the fraction of EM shower energy in a given LYSO:Ce tile within a RADiCAL module schematically shown in \Cref{RadSchem,TType}. The Beam is incident from the left, impacting Layer 0 first. At the bottom of the figure is a schematic of the capillary with the  location of the WLS filament timing element marked. This location within the capillary was optimized for earlier studies at the Fermilab FTBF, where the  electron energy was in the range $20~\leq~E~\leq~30~GeV$ and where shower max occurred in layers 8-10. This position is adequate (although not optimized) for higher energy showers which occur deeper in the module in layers 11-13, and will be corrected in future work, by repositioning and lengthening slightly the WLS filament location.}
    		\label{EDep}
    	\end{center}
    \end{figure}

    For the tests reported in this paper, the RADiCAL module was instrumented with four T-type capillaries as indicated in Fig.~\ref{TType}. The capillaries were of 183~mm length, having an outer diameter of 1150 $\mu$m and an inner (core) diameter of 950 $\mu$m. Within each core,  a DSB1 WLS filament of 900 $\mu$m diameter and 15~mm length was positioned in the region of shower max for EM showers of 25 GeV $\leq$ E $\leq$ 150 GeV. The remainder of each core was filled and fused with quartz rods as described above. The capillaries then transmitted the wave-shifted light to Hamamatsu HDR2 SiPM photosensors which were mounted on readout cards postioned at both the upstream and downstream ends of the module. The central hole in the module (as shown in Fig.~\ref{LysoPlate} was unused in these studies, but is reserved for future use for timing measurement, energy measurement or calibration. A GEANT4 study (Fig.~\ref{EDep}) shows the fraction of EM shower energy deposited in successive LYSO:Ce tiles in the module, indicating that the location of shower max varies slowly with energy, allowing useful placement of wavelength shifting filaments for timing.
    
\section{Expectations from simulation}
\label{EFS}

\begin{figure}
    	\begin{center}
    		\includegraphics[width=0.45\textwidth]{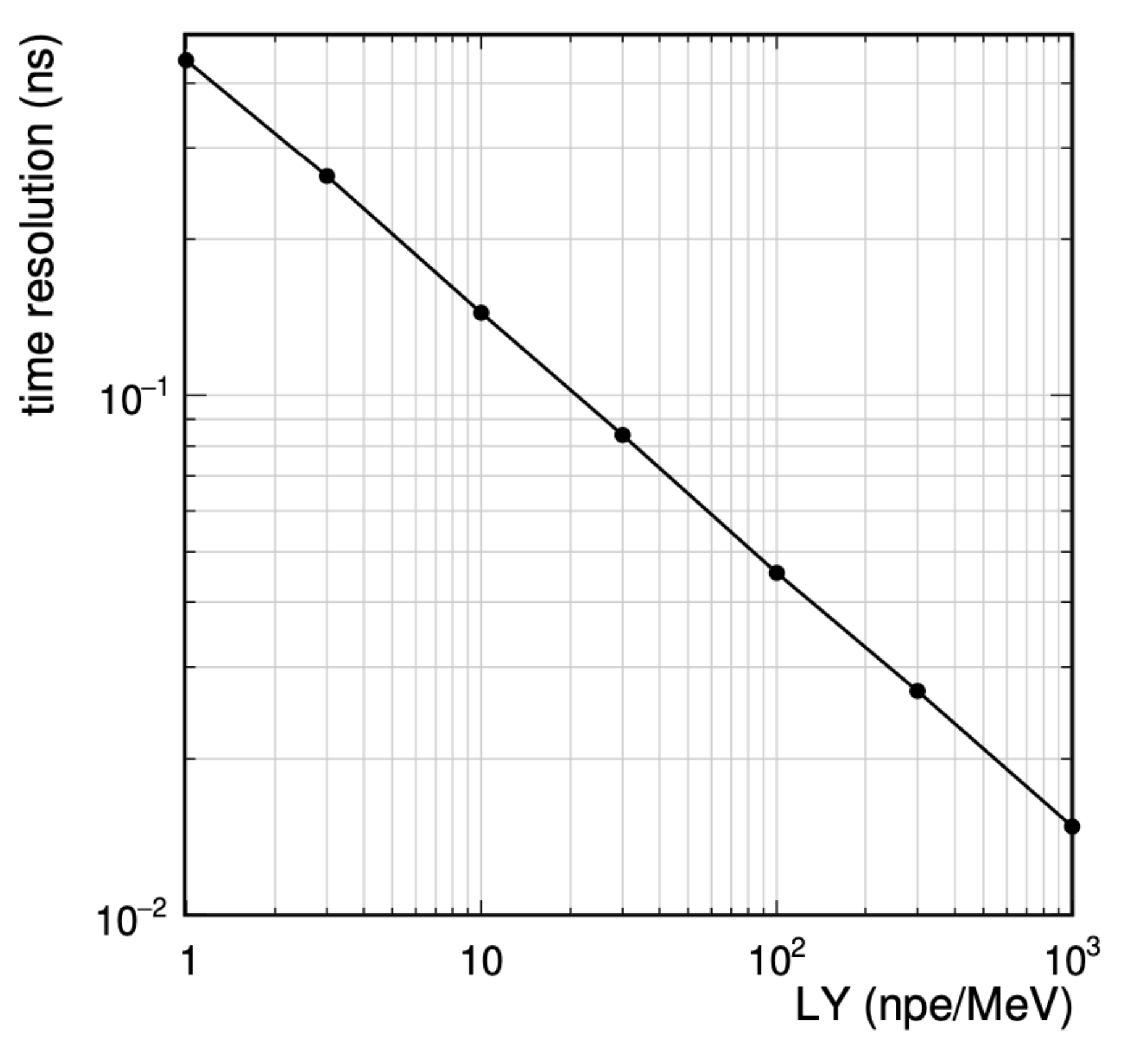}
    		\caption{Timing resolution vs detected light yield in photoelectrons per MeV, simulated for a 50 GeV electron shower. Only the downstream readout was used in this GEANT4 simulation study.}
    		\label{TimingSasha}
    	\end{center}
    \end{figure}

\begin{figure}
    	\begin{center}
    		\includegraphics[width=0.45\textwidth]{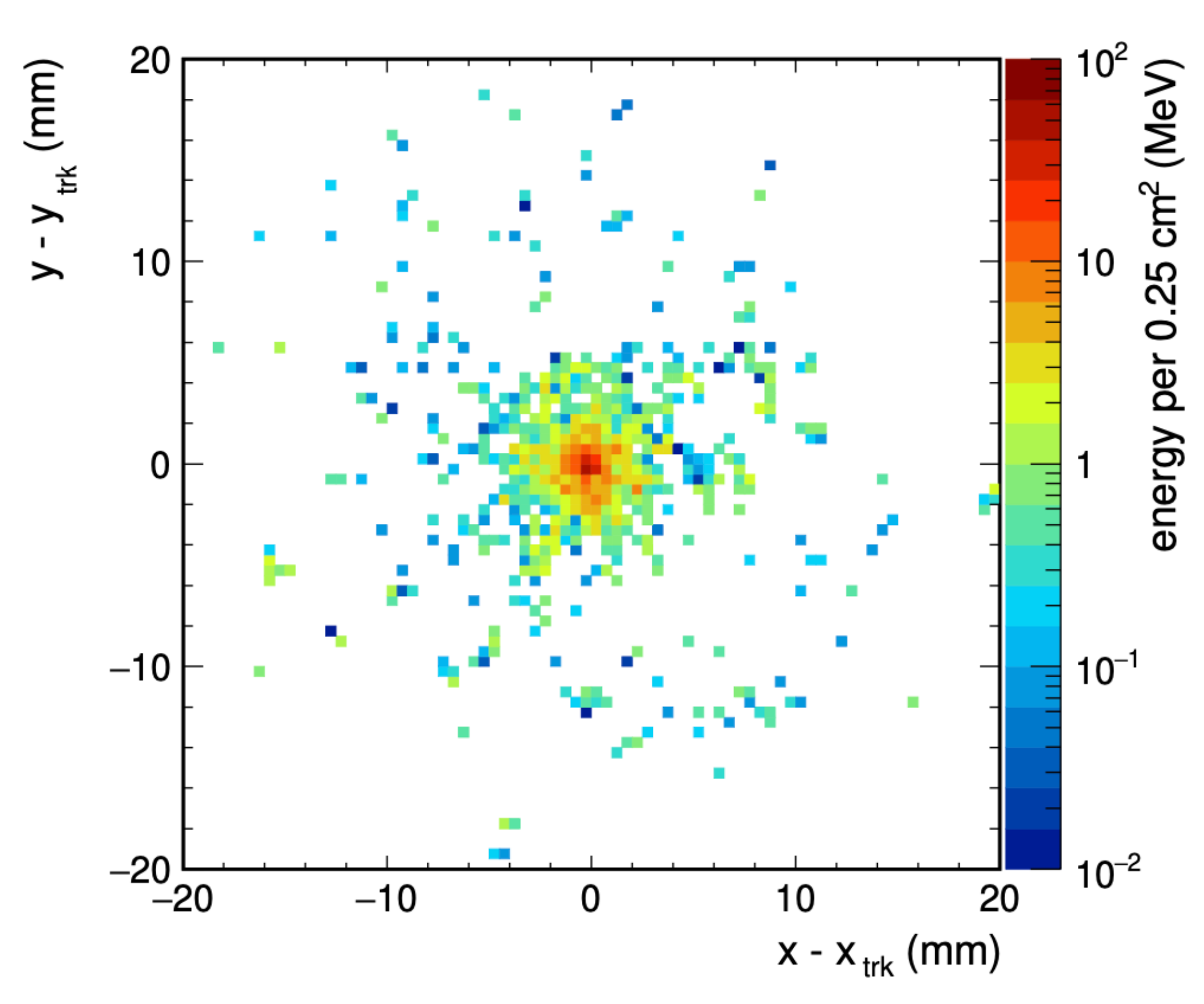}
    		\caption{Location of the energy in a RADiCAL module at shower max for 50 GeV electrons in GEANT4 simulation. The boundary of the module itself is a square of area 14 x 14 mm$^2$ located at the center of the figure.}
    		\label{Moliere}
    	\end{center}
    \end{figure}

    A GEANT4 simulation \cite{ledovskoy2021} was used to assess the potential for precision timing resolution when measured from the shower max region. For this study, the simulation assumed only one SiPM readout at the downstream end of a T-type capillary inserted through the center of the module.  The simulation indicated excellent timing resolution could be achieved dependent upon detected light level, Fig~\ref{TimingSasha}. Also noteworthy was that the shower max timing signal is derived from a region of very small transverse size r $\leq$ 5~mm, Fig.~\ref{Moliere}, with a shower radius at that location, which is significantly smaller than the Moli\`{e}re Radius and more closely approximating the radiation length of the module. This compact region contains one to two orders of magnitude more charged shower particles than would be created by the passage of  a minimum ionizing particle (MIP) in that location, creating a very large and localized optical pulse. The timing resolution is dominated by the measured rise time of this signal. Fig.~\ref{TimingSasha} indicates that, in simulation, a timing resolution of significantly less than 50 ps could be achievable with RADiCAL modules depending upon the detected light level. By reading out the light from both upstream and downstream ends of timing capillaries, and by using several timing capillaries per module, the aim has been to reach or exceed this expectation for the timing resolution of a RADiCAL module. 

    The shower max method also provides an additional pattern recognition constraint. In the modular array, the module that is struck by an incoming electron (or gamma) provides the predominant timing signal for the EM shower. In adjacent modules, the timing signals are effectively absent, since there is very little energy deposited in them in that region, given the very compact transverse size of the shower at the shower max location, Fig.~\ref{Moliere}. This is a potentially powerful locational tool to identify which module is struck by an incoming EM particle in the higher $\eta$ (more forward) regions of collider detectors or fixed target experiments where separation and reconstruction of overlapping showers would be potentially challenging.
    
    These simulations suggested that the shower max timing technique, combined with fast, local energy measurement, would be a potentially powerful technique for measuring both precision timing and shower location for EM particles (electrons, positrons and gammas).

\section{Module assembly and beam test arrangement}
\label{MABTA}

\begin{figure*}[ht]
       \centering
       \begin{subfigure}[b]{0.32\textwidth}
          \includegraphics[width=\linewidth]{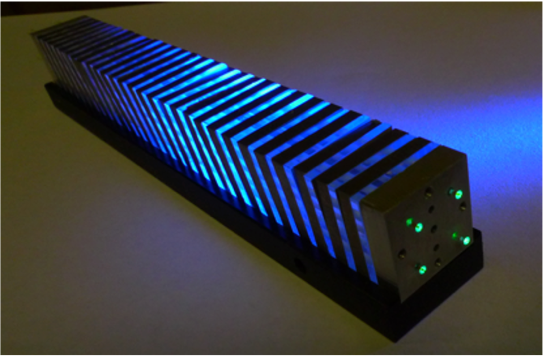}
          \caption{}
          \label{fig:sub1}
       \end{subfigure}
       \hfill
       \begin{subfigure}[b]{0.32\textwidth}
          \includegraphics[width=\linewidth]{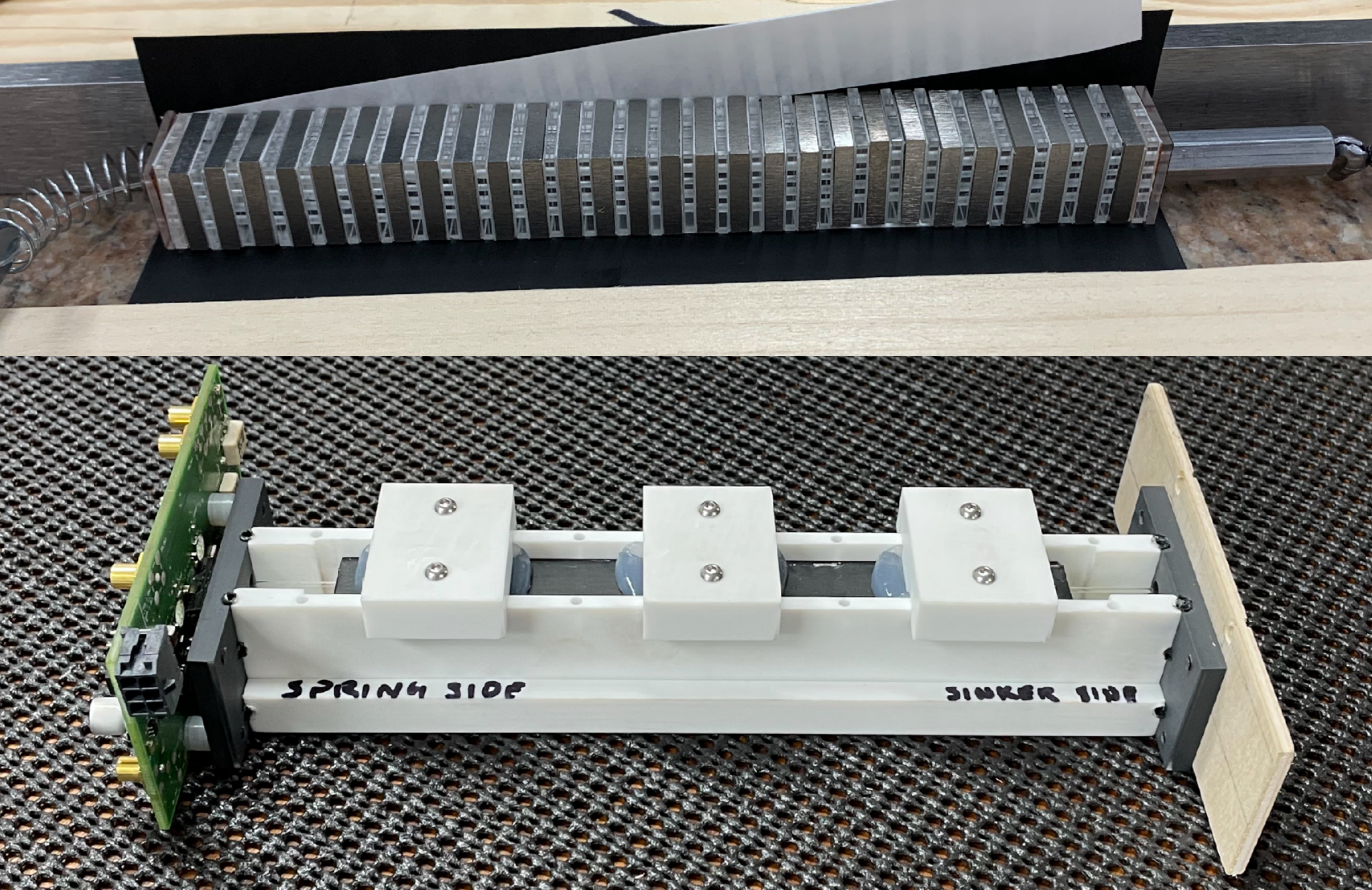}
          \caption{}
          \label{fig:sub2}
       \end{subfigure}
       \hfill
       \begin{subfigure}[b]{0.32\textwidth}
          \includegraphics[width=\linewidth]{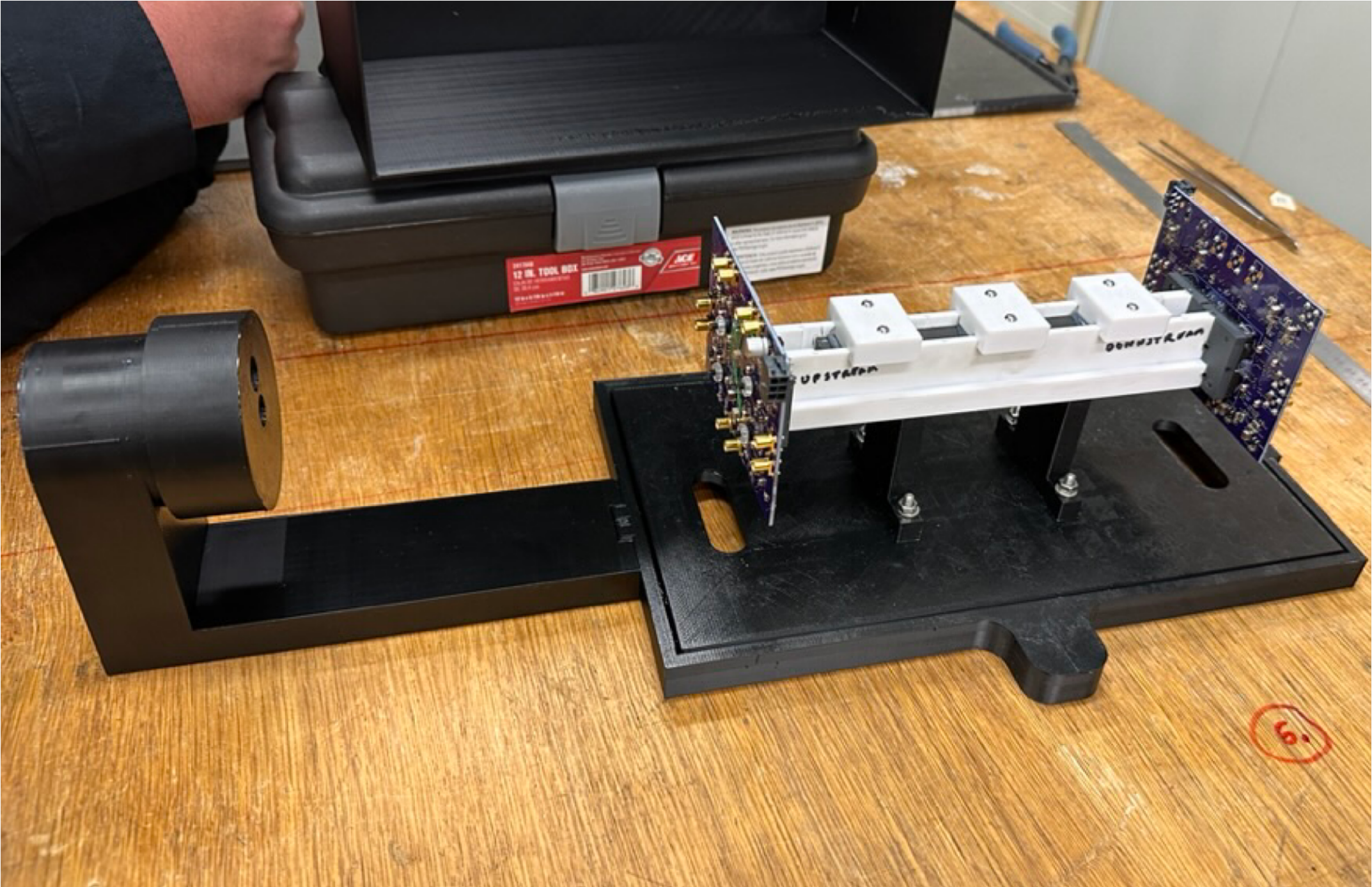}
          \caption{}
          \label{fig:sub3}
       \end{subfigure}
       \caption{The RADiCAL modular structure exposed to UV light at 365 nm, showing fluorescence of LuAG:Ce tiles at 425nm and DSB1 WLS emission at 495nm at the ends of the capillaries, (a). Assembly of a RADiCAL module. The top shows the tiles being stacked; the bottom shows the Delrin assembly, (b). The 3D printed mounting structure that supported the MCP tube, in the cylindrical enclosure on the left, and the RADiCAL module on the right with front end readout cards installed, (c).}
       \label{fig:rad}
    \end{figure*}

\begin{figure*}[ht]
       \centering
       \begin{subfigure}[b]{0.22\textwidth}
          \includegraphics[width=\linewidth]{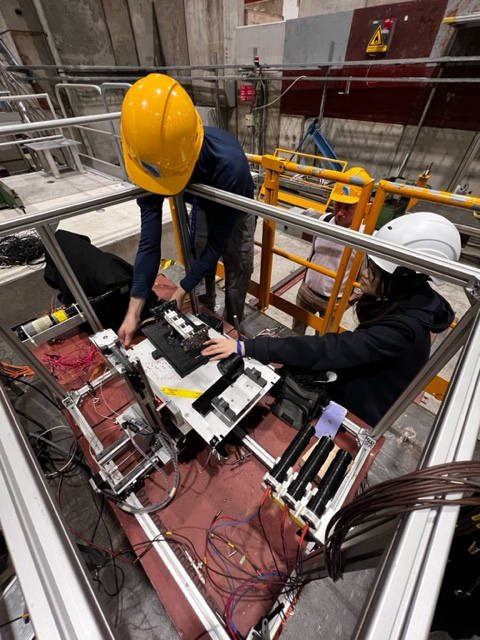}
          \caption{}
          \label{fig:sub1a}
       \end{subfigure}
       \hfill
       \begin{subfigure}[b]{0.352\textwidth}
          \includegraphics[width=\linewidth]{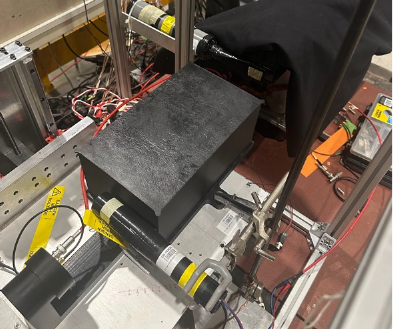}
          \caption{}
          \label{fig:sub2b}
       \end{subfigure}
       \hfill
       \begin{subfigure}[b]{0.32\textwidth}
          \includegraphics[width=\linewidth]{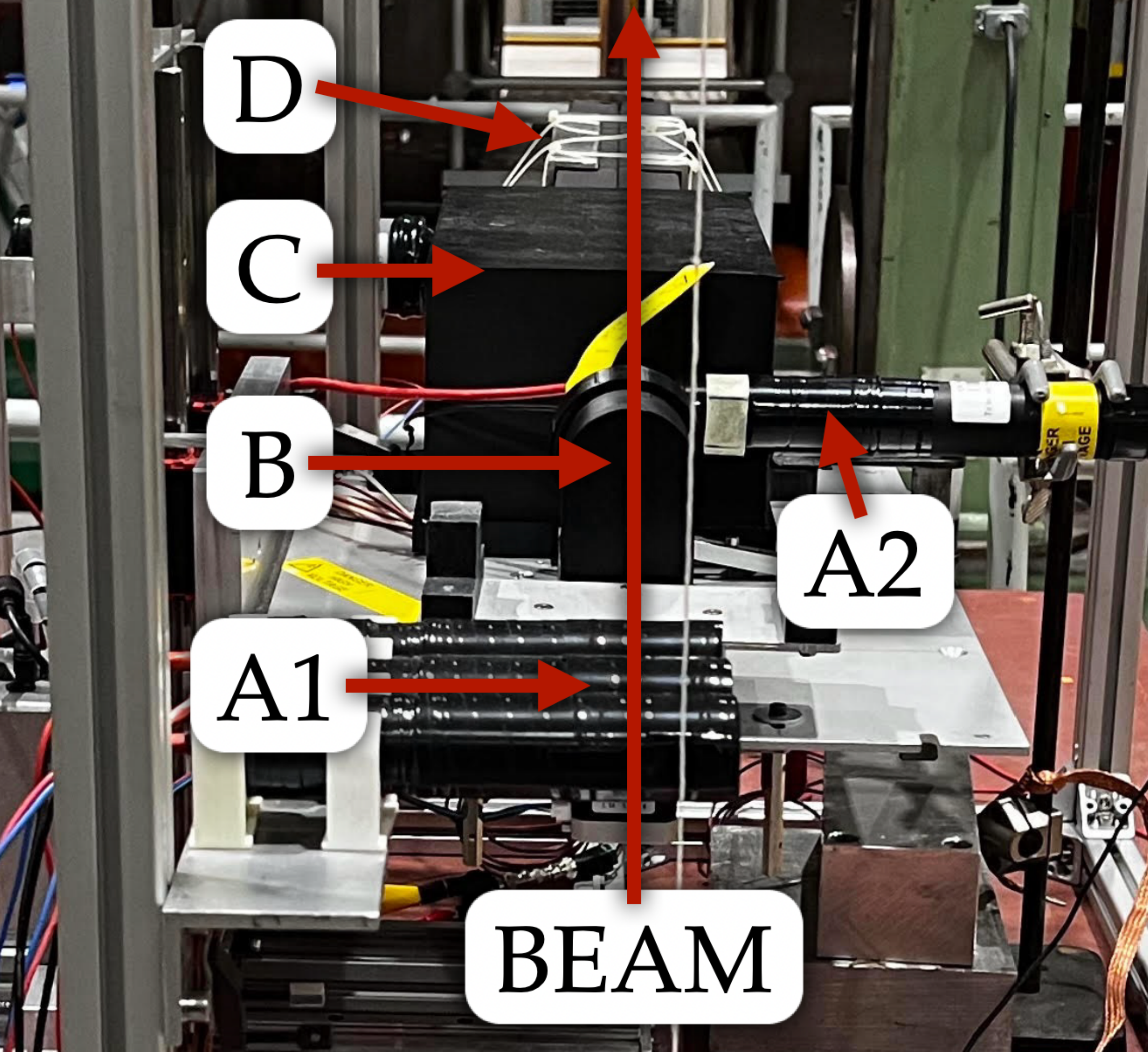}
          \caption{}
          \label{fig:sub3c}
       \end{subfigure}
       \caption{Installation of the RADiCAL module on the CMS/HF Lift Table in the H2 beam line at the North Area Site at the CERN SPS. Beam direction is from lower right to upper left, (a). Overhead closeup view of the relative placement of the MCP tube (lower left), the 2 x 2 cm$^2$ trigger counter, the RADiCAL module underneath the 3D-printed enclosure and the upstream end of the Pb glass array concealed beneath black cloth (upper right). The beam enters this picture from the lower left corner and exits upper right., (b). The 3D printed mounting structure that supported the MCP tube, in the cylindrical enclosure on the left, and the RADiCAL module on the right with front end readout cards installed, (c).}
       \label{fig:setup}
    \end{figure*}

    The RADiCAL module shown schematically in Fig.~\ref{RadSchem} and under UV illumination in Fig.~\ref{fig:rad}a was assembled in a Delrin structure. WLS capillary tubes were then inserted and front end electronics cards supporting the SiPM were attached to the upstream and downstream ends of the structure as shown in Fig.~\ref{fig:rad}b. The module was then mounted on a 3D-printed base plate which held the RADiCAL module and, just upstream, a micro channel plate (MCP) phototube which provided high resolution reference timing, Fig.~\ref{fig:rad}c. Once bias voltage cables and readout cables to CAEN digitizers were installed, the RADiCAL module was covered with a 3D-printed light-tight cover that fit securely into grooves in the base plate. The assembly as then taken to the North Area/Pr\`{e}vessin Site at the CERN SPS, where it was installed on the CMS/HF Motion Table in the H2 beam line (see Fig.~\ref{fig:setup}a and Fig.~\ref{fig:setup}b).   

    The final experimental arrangement of detectors in the beam line is shown in Fig.~\ref{fig:setup}c. Specifically, scintillation counters of dimension 1 x 1 cm$^2$ and 2 x 2 cm$^2$ (labeled A1 and A2 respectively in the figure) provided triggering options. The A2 counter served as the primary trigger for the data taking, since its area fully covered the entire active area of the RADiCAL module. The MCP tube (Hamamatsu R3809U-50) with a timing resolution of 10~ps $< \sigma_{t} < 20$ ps depending on detection region within the tube, and labeled B in the figure, was used as an independent precision timing reference. The RADiCAL module, labeled C, was next in the sequence and was followed by a 2 x 2 array of Pb glass counters, labeled D, forming a detector of 4 x 4 x 40~cm$^3$ which served as a backing calorimeter to measure any shower energy leakage from the RADiCAL module and for identification of any potential hadron (MIP) background in the beam. Not shown in Fig.~\ref{fig:setup}c was a beam chamber provided by the CERN SPS Coordination Team, which provided sub-millimeter horizontal and vertical (x, y) position information of the incoming beam at the location of the RADiCAL module.

\begin{figure*}
        \centering
        \includegraphics[width=0.8\linewidth]{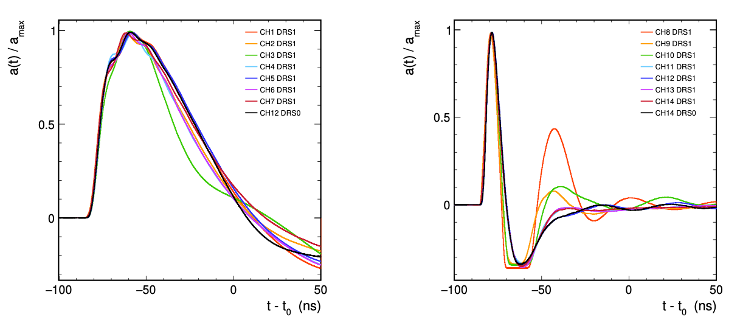}  
        \caption{Average normalized waveforms for signals from the SiPM amplified at low gain (left) and high gain (right). The low gain signals were used for localized energy measurement from the shower max region. The high gain signals were used for fast timing using the fixed threshold on the rise of the leading edge of the pulse.}
        \label{fig:waveforms}
    \end{figure*}

    The custom-built front-end electronics cards each supported four Hamamatsu HDR2 SiPM and supported on-board low-gain and high-gain amplification of the SiPM signals. The low gain amplification was for local energy measurement in the shower max region; the high gain signals were differentially amplified to provide fast rise-time pulses for timing measurement and were modeled after a CERN design \cite{cates2018}. All signals were digitized with CAEN DT5742 digitizers. Fig.~\ref{fig:waveforms} shows characteristic pulse shapes of these signals recorded in an earlier data run at Fermilab \cite{wetzel2023-2} and which were similar for this data. The local shower energy (at Shower max) could be determined by the pulse height or pulse area of the low gain signals, Fig.~\ref{fig:waveforms} Left; the shower timing for each SiPM channel was determined by setting a fixed threshold on the leading edge of the pulse rise of the high gain signals, Fig.~\ref{fig:waveforms} Right. For the data reported here, the high gain signals were driven further into saturation, to maximize the signal rise of the leading edge of the pulses for improved timing response.
    
\section{Module performance}
\label{MP}

    The CERN H2 electron beam was directed to the detector apparatus shown in Fig.~\ref{fig:setup}c, in successive 25 GeV steps in beam energy, covering the range 25 GeV $\leq E_{beam} \leq$ 150 GeV.  At least ${10^6}$ triggers were recorded at each step for data analysis. 
    
    \subsection{Measurement of EM shower energy at shower max}
    \label{MP1}
    
        As shown in Fig.~\ref{EDep} above, simulation studies indicated that a  portion of the total EM shower energy was sampled by $\approx$ 3 LuAG:Ce tiles at shower max, and the WLS filaments positioned in the T-type capillaries at that location provided a wave shifted optical signal from that location which was then transmitted via the capillaries to the SiPM readout. The SiPM signals were then amplified at low-gain for position-based energy measurement in the shower max region and at high-gain, with differential amplification for fast timing measurement. 
    
\begin{figure*}
            \centering
            \includegraphics[width=0.49\linewidth]{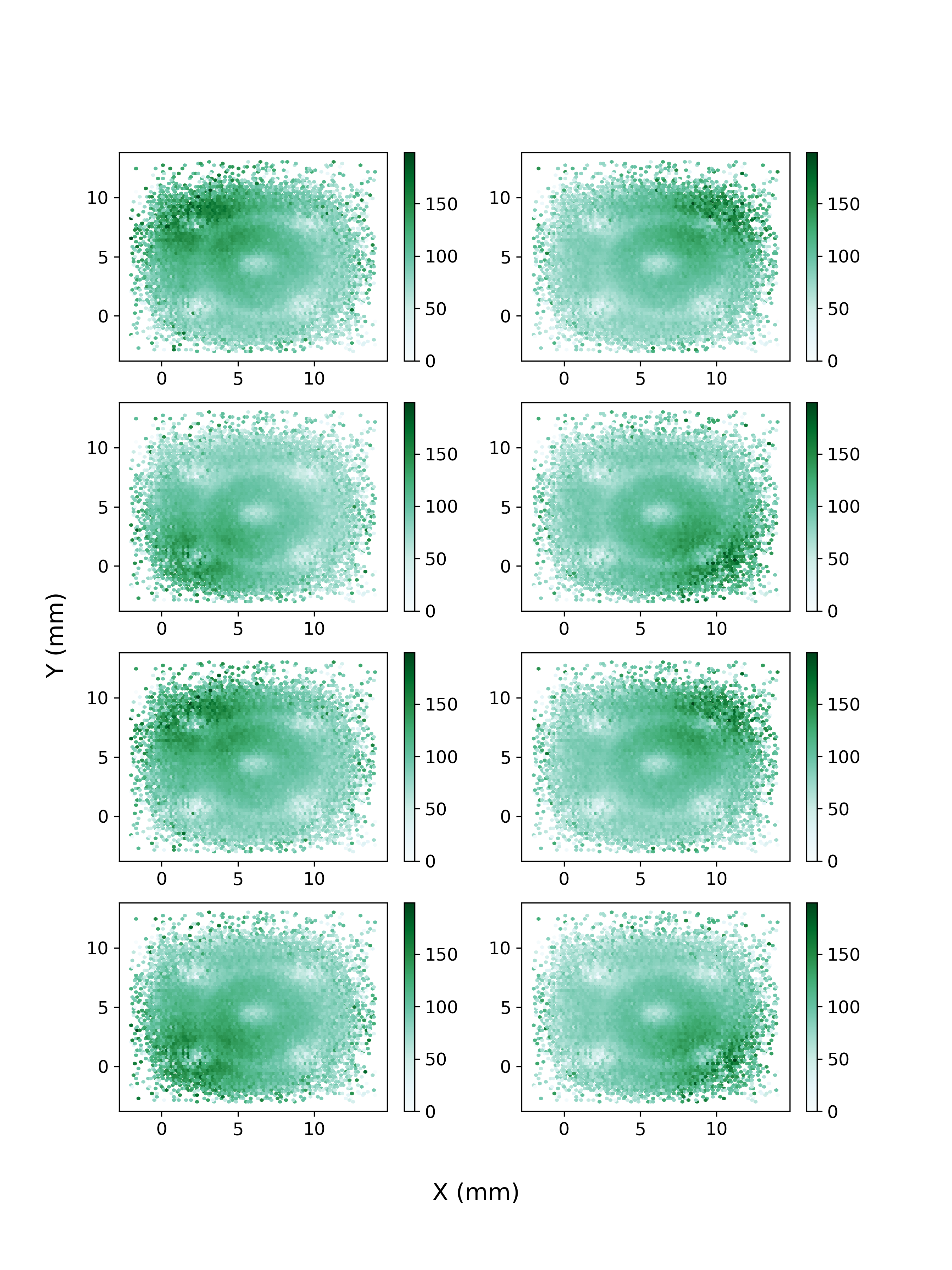}
            \includegraphics[width=0.49\linewidth]{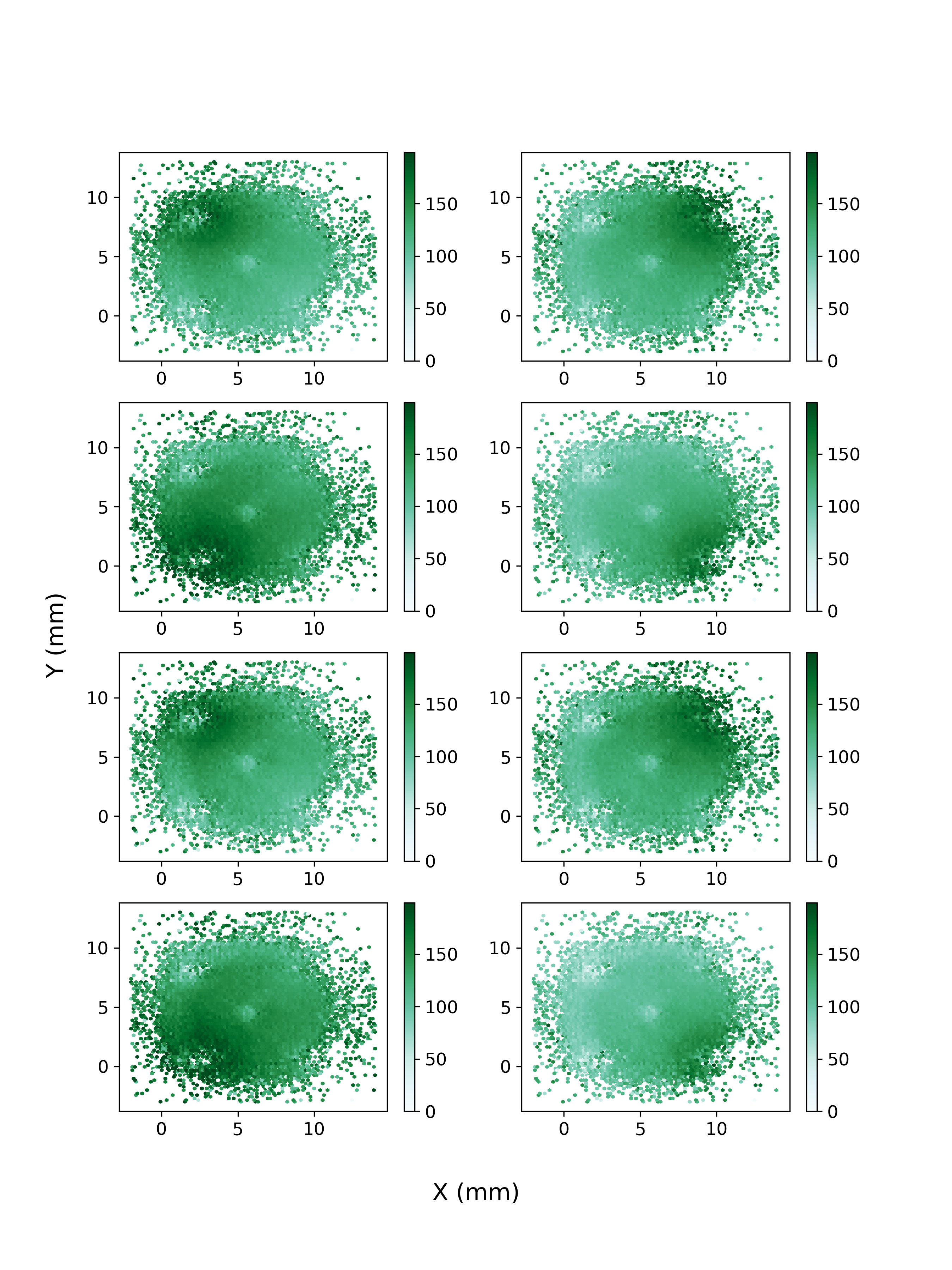}
            \caption{For $E_{beam}$ = 25 GeV:  Beam particle location horizontally and vertically (x, y) at the RADiCAL module provided by the upstream beam chamber, and with the data triggered by the A2 counter. Color density is given by the pulse height in the low-gain amplifiers for each SiPM. Left four plots are for the upstream SiPM and the right four plots are for the downstream SiPM. The capillaries and central (unused hole) are clearly seen. The faint ring is likely due to an internal seal within the MCP. The MCP was located just upstream of the RADiCAL module in the beamline.}
            \label{fig:position}
        \end{figure*}

    
        Fig.~\ref{fig:position} shows the response from each of the capillaries in the module as a function of electron beam (x, y) position as determined from the beam line wire chamber and for two beam energies, 25 GeV and 125 GeV. The responses displayed are based on the amplitudes of the individual low-gain signals from each of the four T-type capillaries and their SiPM readout at both upstream and downstream ends of the module. The locations of the four capillaries and the (uninstrumented) central hole are clearly identifiable, since the WLS material is not scintillating, but rather wave shifting. It is also clear that the closer the electron beam is to a given capillary, the stronger the WLS fiber signal is seen. Also seen in the images at 25 GeV is a faint ring, which is suggestive of an internal seal within the upstream MCP tube, causing some showers to be initiated upstream of the RADiCAL module, reducing the detected energy at shower max. The presence of the ring indicates that the RADiCAL module and the MCP are very well aligned in the beam. 
    
    \subsubsection{Energy measurement in the module for $E_{beam}$~=~25~GeV}
    \label{E25}
\begin{figure*}
            \centering
            \includegraphics[width=0.8\linewidth]{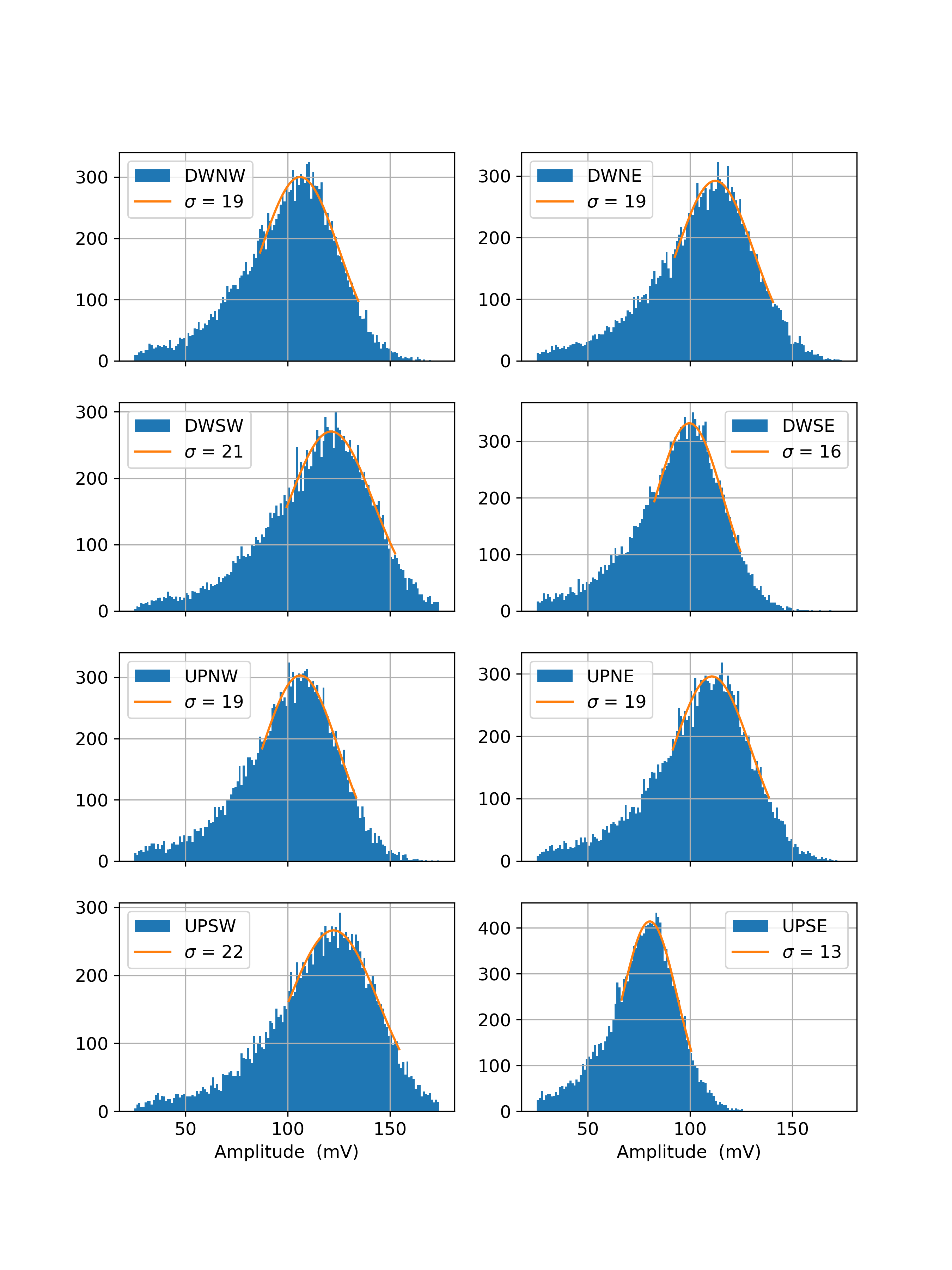}
            \caption{For $E_{beam}$ = 25 GeV: Distribution of the pulse heights from the low gain signals from each SiPM channel.  The four downstream SiPM channels at the top.  The four upstream SiPM channels at the bottom. The SiPM channel labeled UPSE in the lower right had a lower gain than the others.  The Gaussian fits to the peak values of all the channels were subsequently scaled to a common value for local energy and position measurement.}
            \label{fig:pulseheights}
        \end{figure*}

\begin{figure*}
            \includegraphics[width=\linewidth]{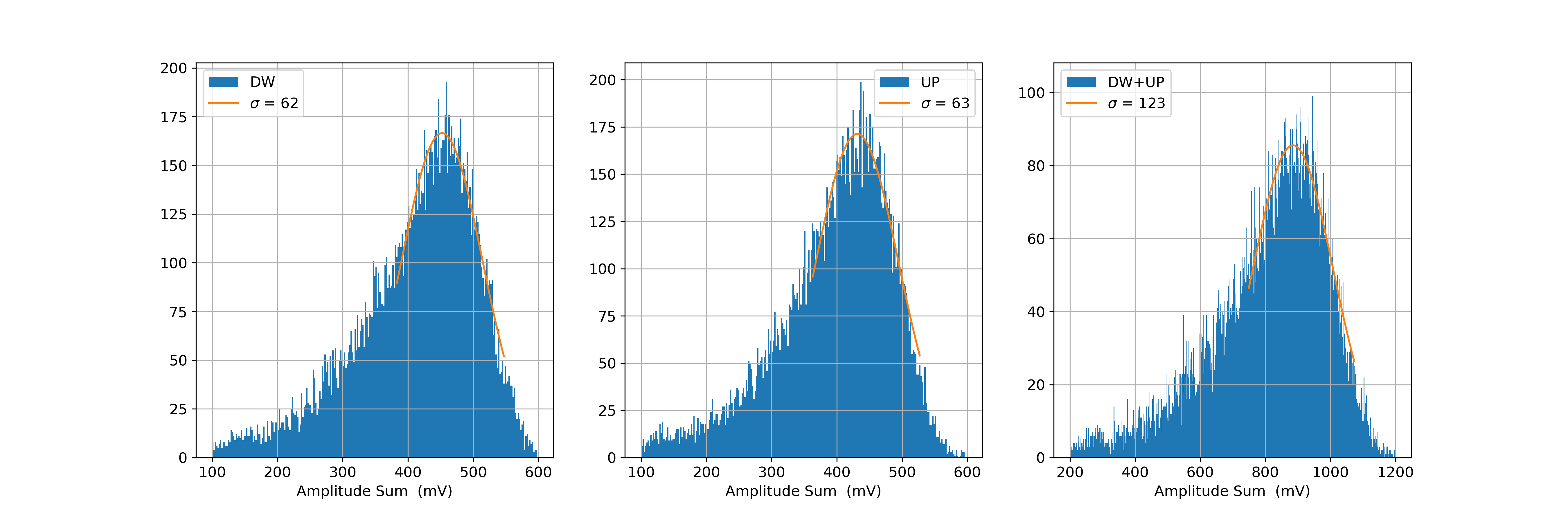}
            \caption{For $E_{\text{beam}}$ = 25 GeV: Sum of the calibrated low-gain signals: Left: for the downstream SiPM channels; Middle: for the upstream channels. Right: the sum of all channels: the peak of the Gaussian fit to this distribution $E_{\text{meas}}$ (in mV) is set to the beam energy $E_{\text{beam}}$ (in GeV) and the width of the Gaussian fit is the measured energy resolution $\sigma_{\mathrm{E_{\text{meas}}}}$.}
            \label{fig:signals}
        \end{figure*}

        Fig.~\ref{fig:pulseheights} shows the distributions of the amplitudes of the low-gain signals for each SiPM channel for a beam energy of 25 GeV. In the analysis, a cut was applied to remove events for which there was a significant response in the downstream Pb glass array. This was to exclude triggered events that either missed the RADiCAL module or were associated with non-EM particles such as hadrons (primarily pions) in the beam. The peaks of these distributions were then scaled to a common value, to adjust for any variation. Then the low gain amplitudes from the downstream SiPM were summed (Fig.~\ref{fig:signals} Left), and separately the the low gain amplitudes from the upstream SiPM were summed (Fig.~\ref{fig:signals} Center). There were no obvious differences in these distributions, and hence downstream and upstream signals were combined (summed) together (Fig.~\ref{fig:signals} Right). Hence the measured energy in the RADiCAL module $E_{meas}$ was given by the summation of the amplitudes of all eight low-gain SiPM signals. This resultant distribution has a significant peak near $E_{meas}$ $\approx{900}$ mV, and a tail that extends down to lower values of measured energy. A Gaussian fit to the peak of the distribution is shown in Fig. 15 Right. The central value of the fit is assigned to the energy of the beam, $E_{beam}$ = 25 GeV, and the standard deviation of the fit is associated with the measured energy resolution in the module at shower max for this beam energy,  $\sigma_{meas}$.  Several factors contribute to the shape of the observed distribution in Fig.~\ref{fig:signals} Right. These include: the particle statistics of the EM showers; shower leakage; photo efficiency and optical coupling of the scintillator and wavelength shifter, and optical transport via the capillaries to the SiPM. These were considered in the simulation discussion above in Section~\ref{EFS} and Fig.~\ref{TimingSasha}. Beam momentum resolution, while present, was but a small contributor and understood to be $\leq$ 2\%, dependent upon collimation and transport settings in the H2 beam line.
    
    \subsubsection{Energy measurement in the module for $25~GeV~\leq~E_{beam}~\leq~150~GeV$}
    \label{Eall}
    
\begin{figure*}
            \includegraphics[width=\linewidth]{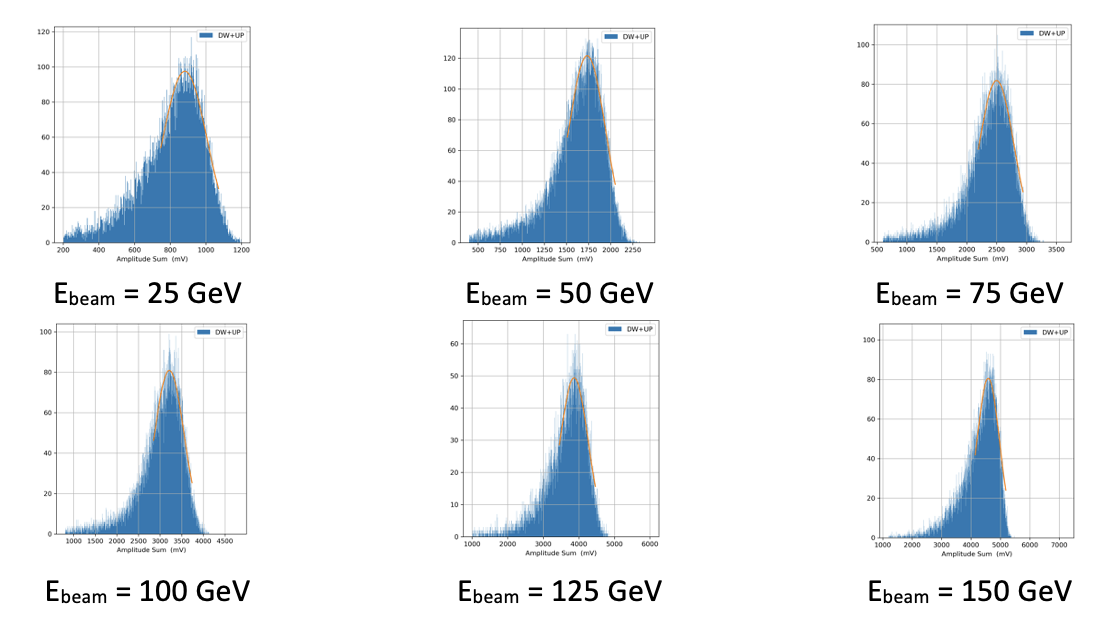}
            \caption{Distributions of summed energy $E_{meas}$ (in mV) from all eight SiPM channels in the RADiCAL module for the six beam energies $E_{beam}$ utilized in this study. The central value of the Gaussian fits (red lines) for each energy (in mV) were then associated with the given nominal energy of the beam (in GeV).}
            \label{fig:nominalenergy}
        \end{figure*}

        The procedure described in Section~\ref{E25} was then applied to the data taken in each successive step in beam energy. The summed amplitudes of all eight low-gain SiPM signals for each triggered event were identified as E$_{meas}$ in mV units for each beam energy step E$_{beam}$ in GeV units from 25 GeV up to 150 GeV. These distributions are shown in Fig.~\ref{fig:nominalenergy}. Notable features in each distribution are a strong peak region which narrows with increasing beam energy, and tails in each distribution to lower measured energy. Gaussian fits were made to each peak region, with the central values of the Gaussians (in mV) associated with the values of the delivered beam energy (in GeV). The standard deviation of each Gaussian was identified as the corresponding energy resolution $\sigma_{meas}$ for that measured energy.

    \subsubsection{Overall behavior of energy measurement at shower max}
    \label{OBEMSM}
    
\begin{figure*}
            \includegraphics[width=0.49\linewidth]{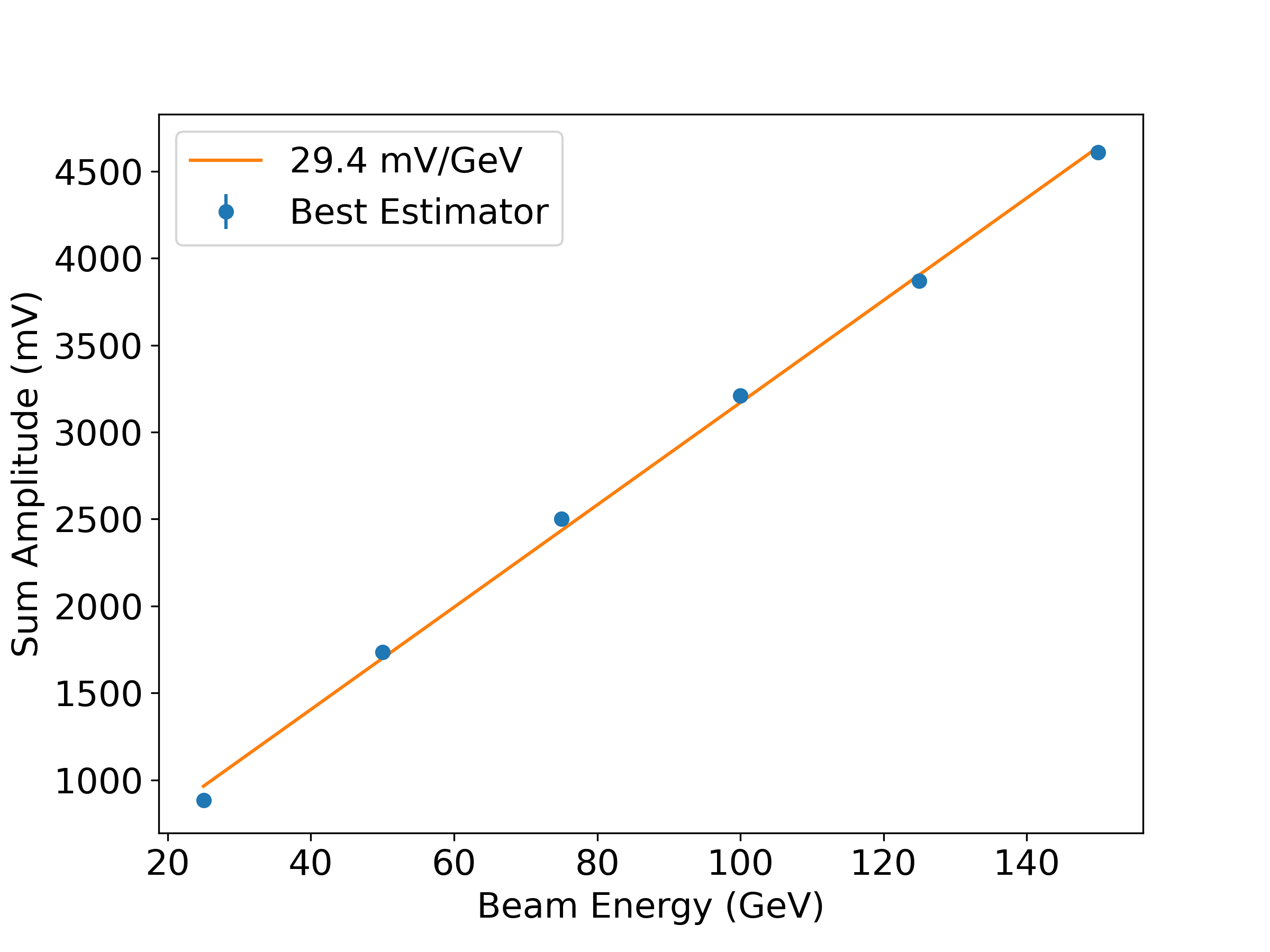}
            \includegraphics[width=0.49\linewidth]{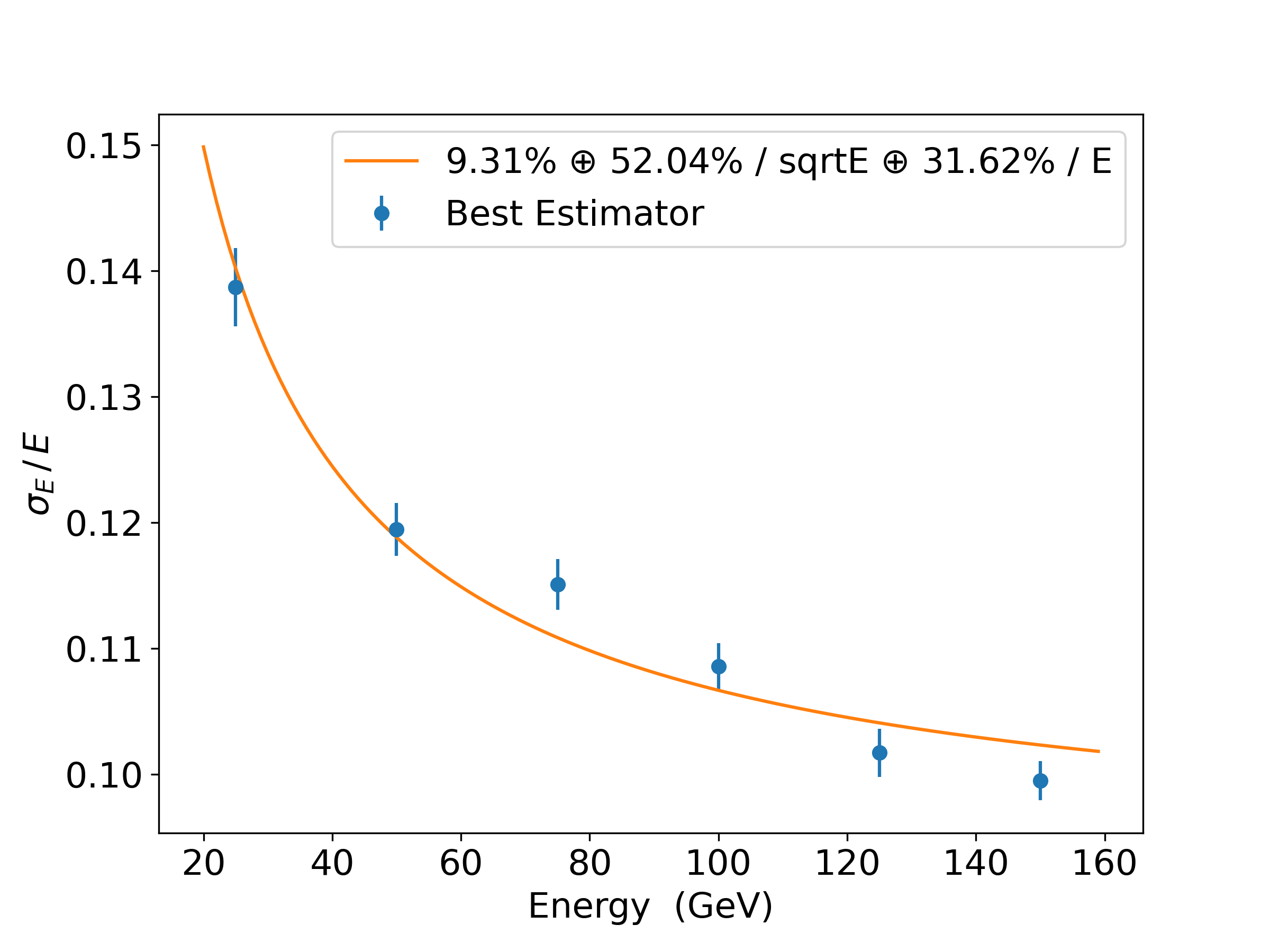} 
            \caption{Left: Plot of the summed energy from all SiPM channels $E_{meas}$ (in mV) vs the nominal beam energy $E_{beam}$ (in GeV), showing a linear slope.  Right: Plot of the measured ratio $\sigma_{E_{meas}}/E_{meas}$ from the shower max region as a function of the nominal beam energy.}
            \label{fig:enResolution}
        \end{figure*}

        For each step in beam energy $E_{beam}$ the central value of the Gaussian fit to each measured energy peak $E_{meas}$ (in mV) was plotted versus the nominal beam energy $E_{beam}$ (in GeV) as shown in Fig.~\ref{fig:enResolution} Left. Linear behavior was observed, with a fitted slope of: $E_{meas}/E_{beam} = 29.4$ mV/GeV. Then in Fig.~\ref{fig:enResolution} Right, for each step in beam energy $E_{beam}$, each measured resolution $\sigma_{E_{meas}}$ (in mV) divided by the central value of the corresponding Gaussian peak $E_{meas}$ (in mV) was plotted as a function of the delivered beam energy. The fit to the data follows the functional form:
    
        \begin{equation}
        	\frac{\sigma_{E_{meas}}}{E_{meas}} = \frac{52.04\%}{\sqrt{E_{beam}}}\oplus\frac{31.62\%}{E_{beam}}\oplus 9.31\%
        	\label{bigequation}
        \end{equation}
        
        Note that this does not represent the energy resolution that would result from a high-resolution measurement of the energy of an EM shower in which all 29 LYSO:Ce layers in the full module would contribute to the measurement, as well as energy from surrounding modules in a 3 x 3 or larger array. Rather, the measurement here corresponds to the energy and energy resolution derived from the restricted region  of shower max, corresponding to sampling over $\approx$ 3 LYSO:Ce tiles within a single module. The key value of this energy measurement is for its importance for the determination of EM shower position localization and for the correlation of the timing resolution with the detected energy at shower max. The former topic, spatial localization, will be discussed in a future paper. The latter topic, correlation of timing resolution with detected energy, is the subject of this paper and elaborated below.
    
    \subsection{Measurement of shower timing.}
    \label{MSTEM}
    
        To determine the signal timing for each SiPM channel, fixed thresholds were set for the high-gain signals for the upstream and downstream SiPM. When the leading edge of the high-gain pulses from  a given channel exceeded the threshold, the timing for that channel was determined and compared with the reference timing provided by the MCP tube.  This allowed for confirmation of the synchronization of arrival times the timing signals, and allowed for averaging of the timing signals.
    
    \subsubsection{General Characteristics of Timing Measurement at $E_{beam}$ = 25 GeV}
    \label{E25Timing}
\begin{figure}
            \includegraphics[width=\linewidth]{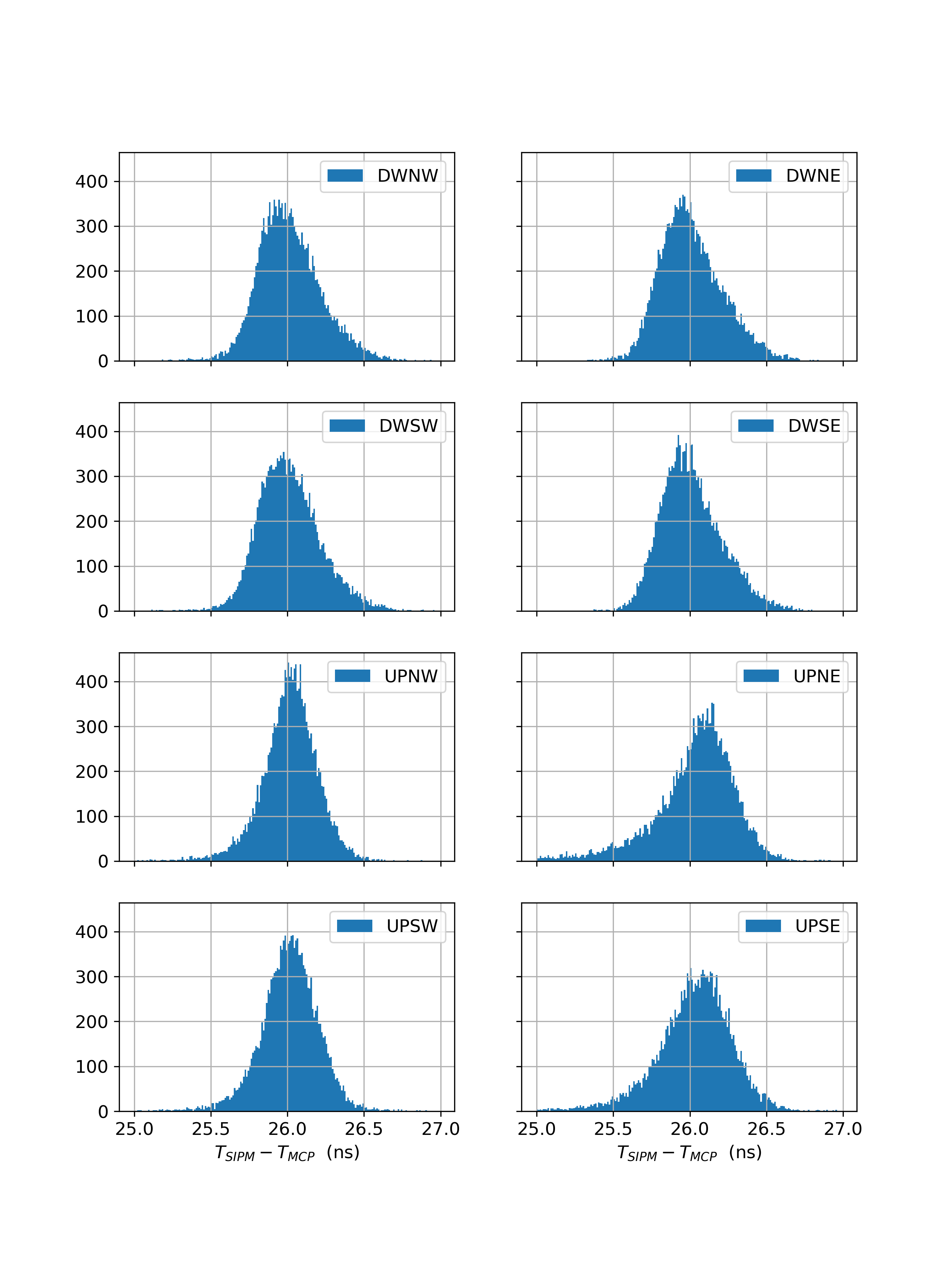}
           \caption{E\textsubscript{beam} = 25 GeV. The timing distributions for each SiPM relative to the MCP timing: $\Delta t$ (i) = (t\textsubscript{SiPM (i)} - t\textsubscript{MCP}). Left: for the four downstream SiPM. Right for the four upstream SiPM.}
           \label{fig:timingdist}
        \end{figure}

\begin{figure}
           \includegraphics[width=\linewidth]{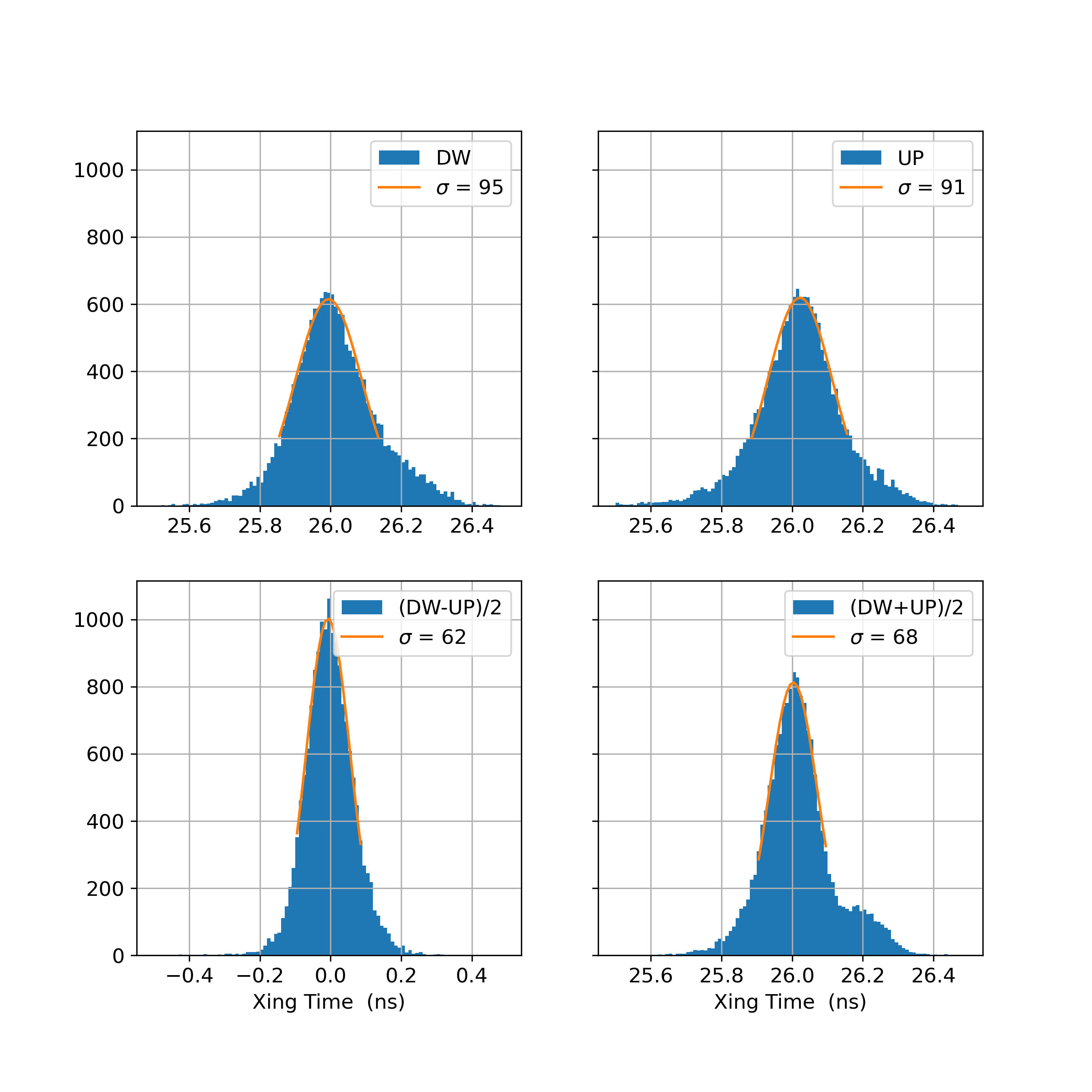}
            \caption{E\textsubscript{beam} = 25 GeV. Upper Left: The average of the timing distributions for the four downstream SiPM. Upper Right: The average of the timing distributions for the four upstream SiPM. Lower Left: The average of the difference of downstream and upstream timing distributions. Lower Right: The average of downstream and upstream timing distributions. Note in this plot the appearance a small satellite peak, which has an area roughly 15\% of the larger peak. Given its small size, its presence has little effect on the timing resolution determined from this distribution (See text)}
            \label{fig:avgtiming}
        \end{figure}

        In Fig.~\ref{fig:timingdist} are shown the timing differences between each of the eight SiPM channels compared with the MCP timing: $\Delta t(i)~=~(t_{SiPM~(i)}~-~t_{MCP})$ for (i=1,8). These timing differences are then averaged separately over the four downstream SiPM channels (i = 1,4): $\Delta t_{DW} = \Sigma~\Delta t(i)/4$, and over the four upstream SiPM channels (j = 5,8): $\Delta t_{UP} = \Sigma~\Delta t(j)/4$. These distributions are shown in Fig.~\ref{fig:avgtiming} in the Upper Left and Upper Right hand plots. Fig.~\ref{fig:avgtiming} Lower Left shows the average difference event by event for these downstream and upstream timing averages: $(\Delta t_{DW}~-~\Delta t_{UP})/2$, and Fig.~\ref{fig:avgtiming} Lower Right shows the overall average of both upstream and downstream timings:  $(\Delta t_{DN}~+~\Delta t_{UP})/2$.  The timing resolution of the RADiCAL module can be determined from Gaussian fits to either of these distributions. Notable in Fig.~\ref{fig:avgtiming} Lower Right is a small amplitude satellite bump whose centroid is shifted by $\approx~0.2~ns$ relative to the main peak. This small peak corresponds to triggered events for which the timing algorithm, which uses a fixed threshold, is shifted due to fluctuation by one sample segment in the CAEN digitzer that is used to digitize the signal timing. This condition affected roughly 15\% of the timing signals of all SiPM channels, causing the shift of $\approx~0.2~ns$ relative to the MCP reference time. The effect appears only the average of the sum of all channels relative to the MCP. It does not appear in Fig.~\ref{fig:avgtiming} Lower Left, as this is the timing difference between downstream and upstream SiPM channels, which is independent of the MCP timing. In any case, the timing resolution is determined by Gaussian fit to the main peaks, and hence the calculation of the timing resolution calculation in the Lower Right hand plot is generally unaffected by the presence of this satellite peak. The main peak in Fig.~\ref{fig:avgtiming} Lower Right is more broad, but this is due primarily to the presence and contribution of the MCP resolution to this distribution, rather than from a contribution from the satellite peak. 
    
\begin{figure*}
            \includegraphics[width=\linewidth]{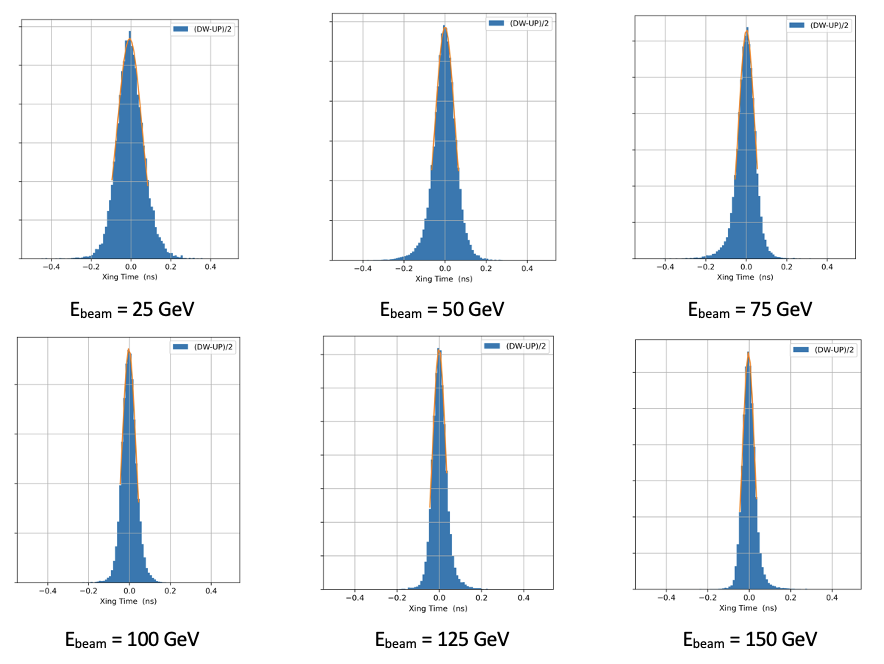}
            \caption{Timing differences between the average times of the four downstream SiPM and the four upstream SiPM in successive steps in beam energy over the range: 25 GeV $\leq$ E $\leq$ 150 GeV.  The Gaussian fits to these distributions provide a preliminary measure of the timing resolution of the module.  The peaks narrow with energy, indicating improvement in timing resolution.}
            \label{fig:timingdiffs}
        \end{figure*}

        The timing resolution, in a generalized way over the energy range 25 GeV$\leq$~E$_{beam}$~$\leq$150~GeV, could then be assessed by taking the average timing of all eight timing differences with respect to the MCP timing: $(\Delta t_{DW}~+~\Delta t_{UP})/2$; or by taking the average of the difference in the time average of the four downstream high gain signals and the time average of the four upstream high gain signals: $(\Delta t_{DW}~-~\Delta t_{UP})/2$. The second method, through subtraction, gives a result which is independent of the MCP timing. Fig.~\ref{fig:timingdiffs} displays the results of the second method for each step in beam energy. The Gaussian fits to the peaks indicate steady improvement in the time resolution as a function of beam energy.
    
        
        
    \subsection{Refined timing analysis}
    \label{RFT}
        What the above overview neglects to consider is that the resolution of the timing signals depends on the detected/measured energy ($E_{meas}$) in the RADiCAL module event by event for a given beam energy ($E_{beam}$), rather than simply given by a unique and fixed energy value for all events. This consideration is based upon earlier study of the timing resolution of the RADiCAL module by our group at the Fermilab Test Beam Facility for which a negative beam of $E_{beam}$ = 28 GeV was available [9], and where a strong dependence was observed for the timing resolution as a function of the detected energy in the RADiCAL module (in mV).
    
        This issue is illustrated by examination of the measured energy signals shown in Fig.~\ref{fig:nominalenergy}, where each of the distributions exhibits a low energy tail. For example, in the case of $E_{beam}$ = 25 GeV, the detected energy at the peak of the distribution is $E_{meas}~\approx$~900 mV, corresponding to the delivered beam energy value, whereas the low energy tail extends down to $E_{meas}~\approx$~200 mV, which corresponds to an observed/detected energy of $E_{meas}$ = 6.7 GeV, based upon the slope of the fit in Fig.\ref{fig:enResolution} Left, relating measured energy $E_{meas}$ (in mV) to Energy (in GeV). To account for such variation, the timing resolution for a given delivered beam energy $E_{beam}$ was therefore studied in bins of measured energy $E_{meas}$ across the peak regions of the energy distributions shown in Fig.~\ref{fig:nominalenergy}.
    
\begin{figure*}
        \centering
        \includegraphics[width=0.9\linewidth]{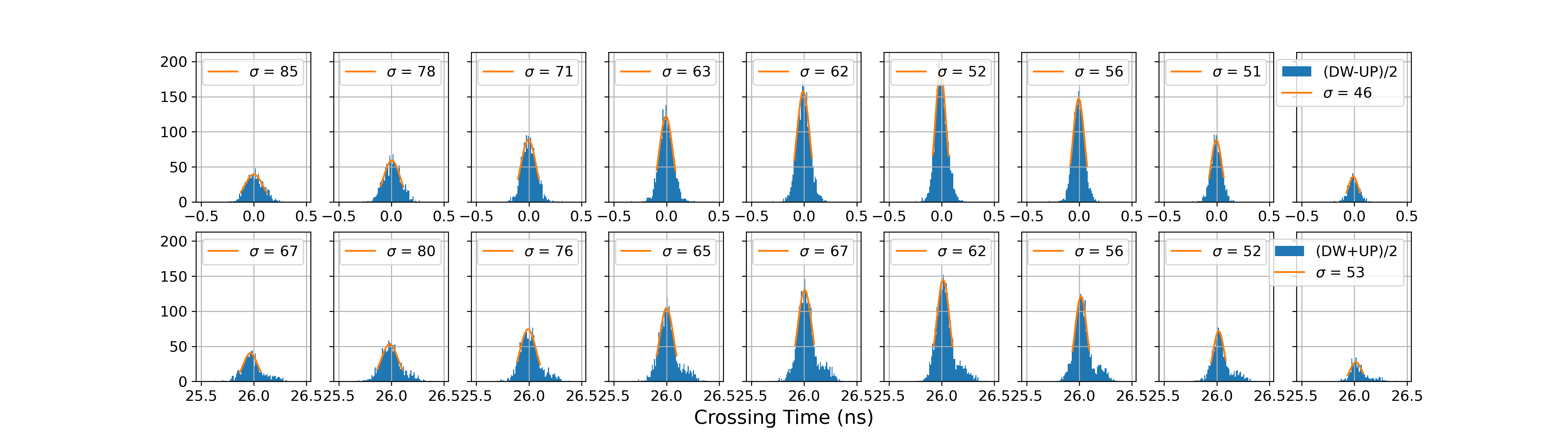}
        \includegraphics[width=\linewidth]{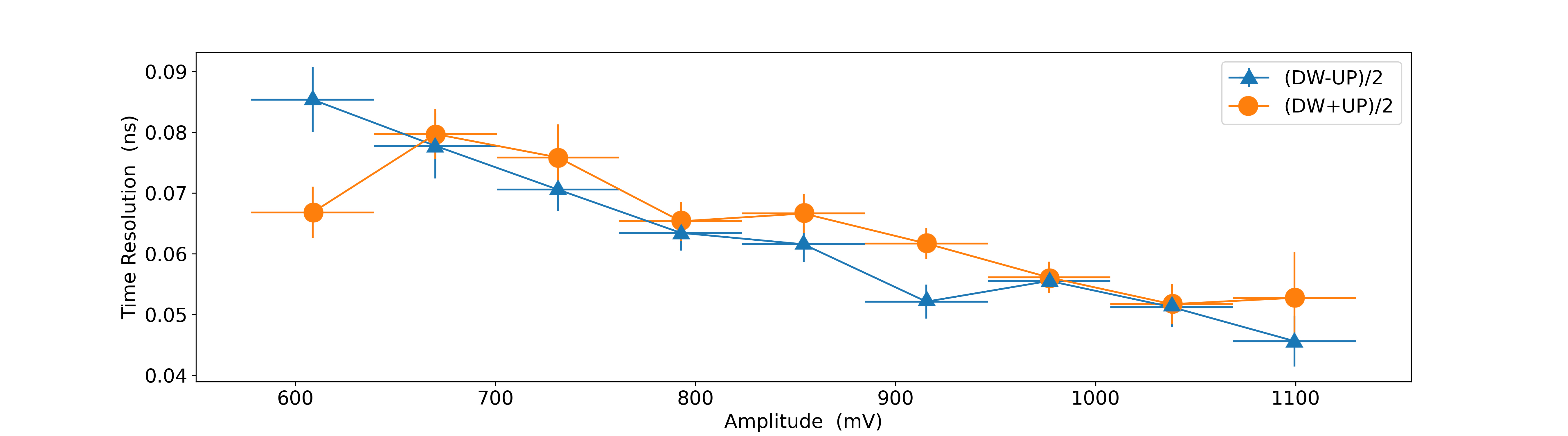}
        \caption{For E\textsubscript{beam} = 25 GeV. The measured energy E\textsubscript{meas} (in mV) has been subdivided into 9 bins, bin 1 at the left, bin 9 at the right in the lower figure. Top: the distributions of $(\Delta t_{DW} - \Delta t_{UP})/2$  for each bin of observed energy E\textsubscript{obs }in mV as indicated by the horizontal error bars in the lower plot. Middle: the distributions of $(\Delta t_{DW} + \Delta t_{UP})/2$ for each bin of observed energy. Lower: the measured time resolution $\sigma_t$ in ps for each bin of observed energy obtained from the standard deviation of the Gaussian fit to the corresponding timing peaks above in the figure. Blue values correspond to $(\Delta t_{DW} - \Delta t_{UP})/2$ and orange values correspond to $(\Delta t_{DW} - \Delta t_{UP})/2$.}
        \label{fig:25GeV}
    \end{figure*}

\begin{figure*}
        \centering
        \includegraphics[width=0.9\linewidth]{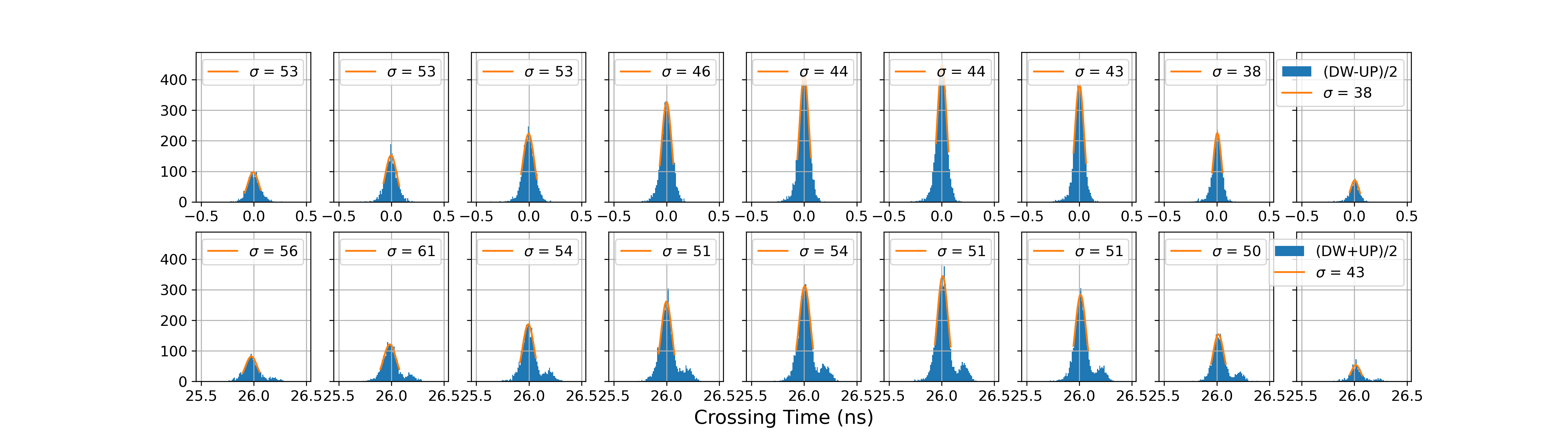}
        \includegraphics[width=\linewidth]{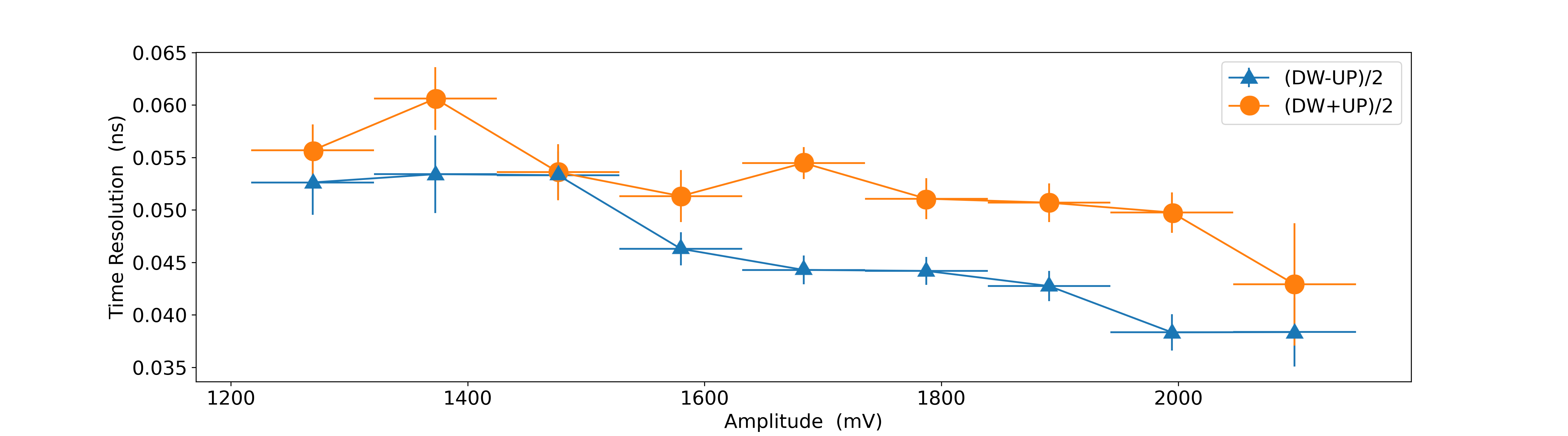}
        \caption{For E\textsubscript{beam} = 50 GeV. The measured energy E\textsubscript{meas} (in mV) has been subdivided into 9 bins, bin 1 at the left, bin 9 at the right in the lower figure. Top: the distributions of $(\Delta t_{DW} - \Delta t_{UP})/2$  for each bin of observed energy E\textsubscript{obs }in mV as indicated by the horizontal error bars in the lower plot. Middle: the distributions of $(\Delta t_{DW} + \Delta t_{UP})/2$ for each bin of observed energy. Lower: the measured time resolution $\sigma_t$ in ps for each bin of observed energy obtained from the standard deviation of the Gaussian fit to the corresponding timing peaks above in the figure. Blue values correspond to $(\Delta t_{DW} - \Delta t_{UP})/2$ and orange values correspond to $(\Delta t_{DW} - \Delta t_{UP})/2$.}
        \label{fig:50GeV}
    \end{figure*}

\begin{figure*}
        \centering
        \includegraphics[width=0.9\linewidth]{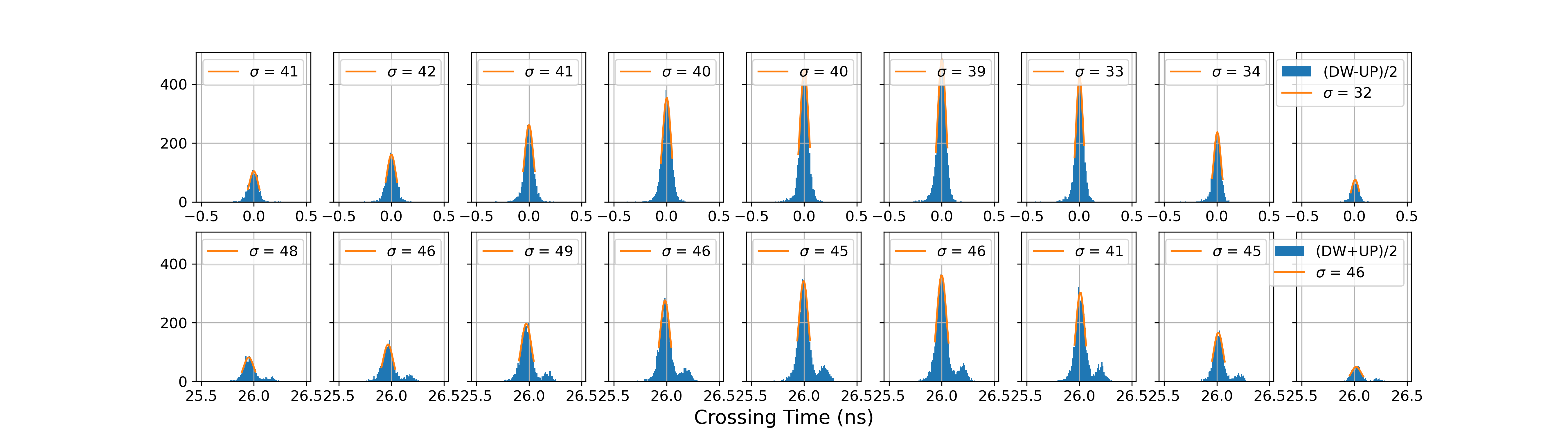}
        \includegraphics[width=\linewidth]{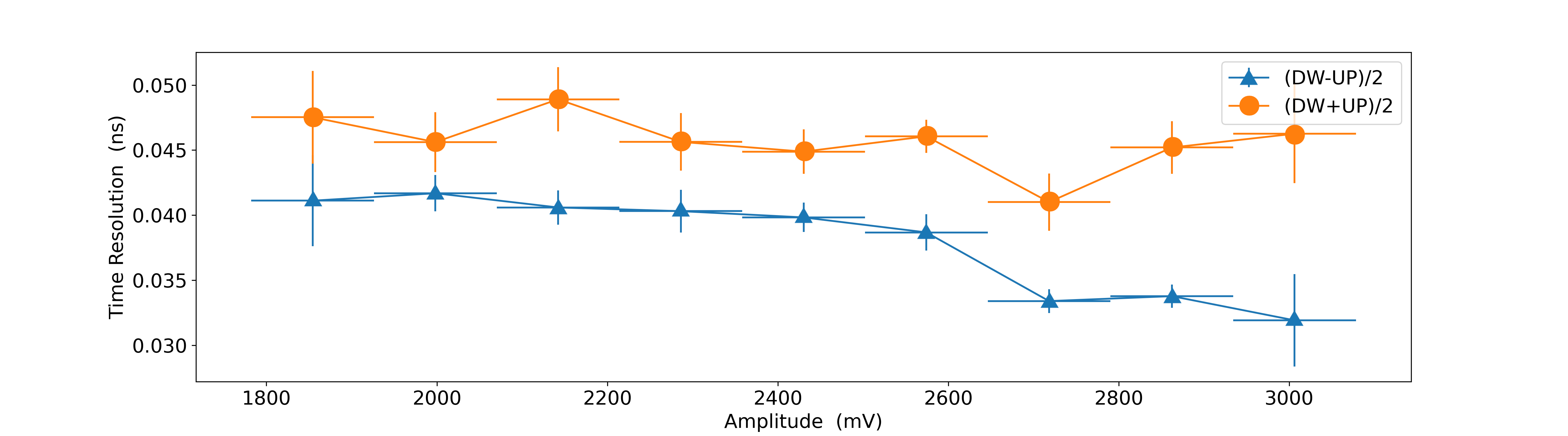}
        \caption{For E\textsubscript{beam} = 75 GeV. The measured energy E\textsubscript{meas} (in mV) has been subdivided into 9 bins, bin 1 at the left, bin 9 at the right in the lower figure. Top: the distributions of $(\Delta t_{DW} - \Delta t_{UP})/2$  for each bin of observed energy E\textsubscript{obs }in mV as indicated by the horizontal error bars in the lower plot. Middle: the distributions of $(\Delta t_{DW} + \Delta t_{UP})/2$ for each bin of observed energy. Lower: the measured time resolution $\sigma_t$ in ps for each bin of observed energy obtained from the standard deviation of the Gaussian fit to the corresponding timing peaks above in the figure. Blue values correspond to $(\Delta t_{DW} - \Delta t_{UP})/2$ and orange values correspond to $(\Delta t_{DW} - \Delta t_{UP})/2$.}
        \label{fig:75GeV}
    \end{figure*}

\begin{figure*}
        \centering
        \includegraphics[width=0.9\linewidth]{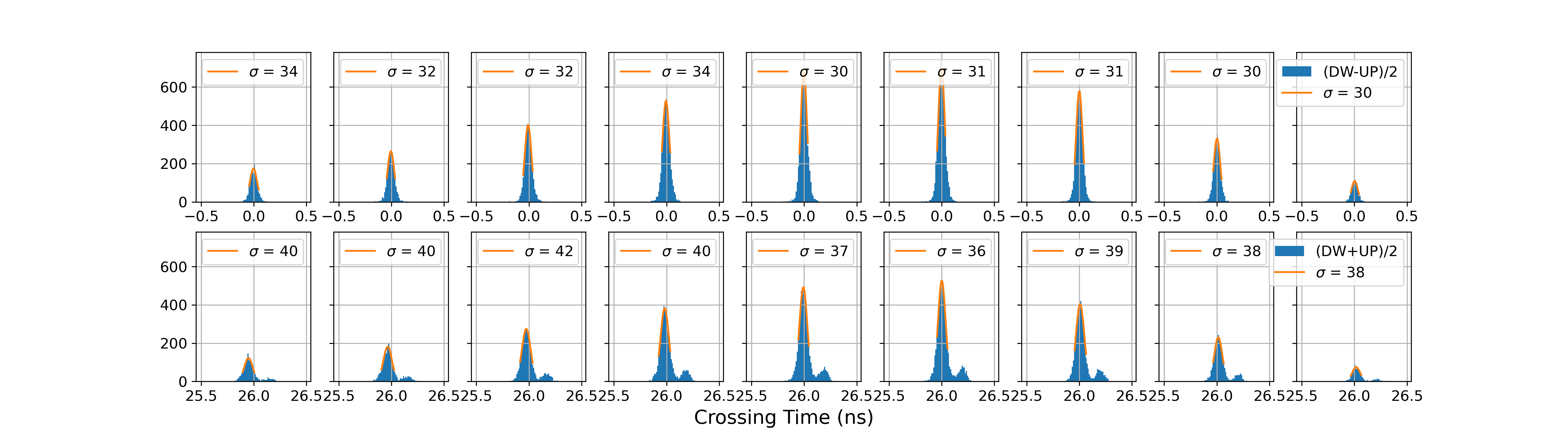}
        \includegraphics[width=\linewidth]{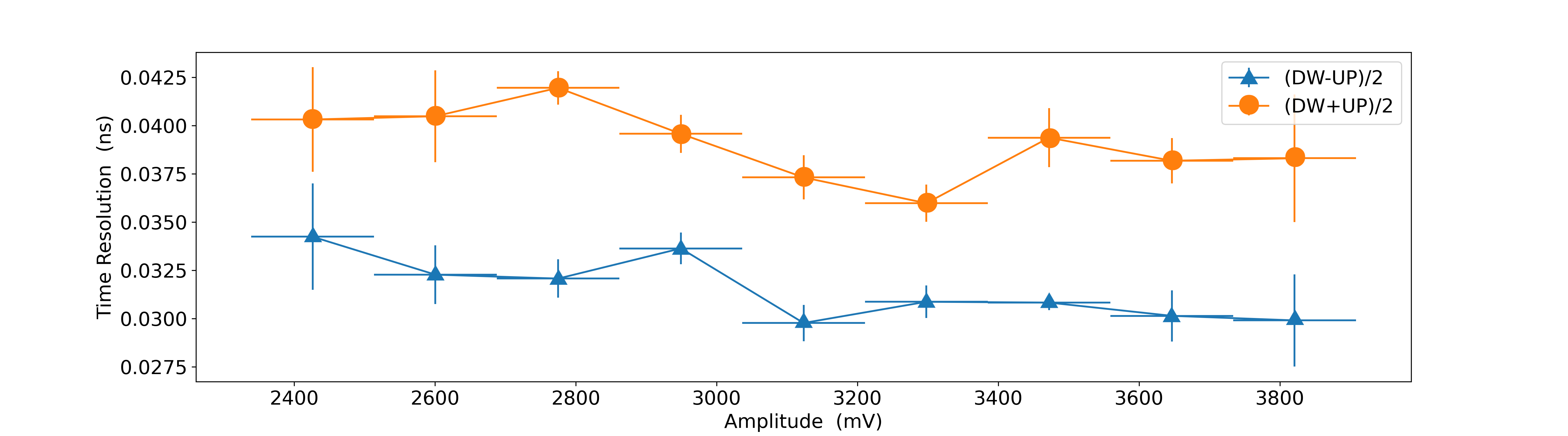}
        \caption{For E\textsubscript{beam} = 100 GeV. The measured energy E\textsubscript{meas} (in mV) has been subdivided into 9 bins, bin 1 at the left, bin 9 at the right in the lower figure. Top: the distributions of $(\Delta t_{DW} - \Delta t_{UP})/2$  for each bin of observed energy E\textsubscript{obs }in mV as indicated by the horizontal error bars in the lower plot. Middle: the distributions of $(\Delta t_{DW} + \Delta t_{UP})/2$ for each bin of observed energy. Lower: the measured time resolution $\sigma_t$ in ps for each bin of observed energy obtained from the standard deviation of the Gaussian fit to the corresponding timing peaks above in the figure. Blue values correspond to $(\Delta t_{DW} - \Delta t_{UP})/2$ and orange values correspond to $(\Delta t_{DW} - \Delta t_{UP})/2$.}
        \label{fig:100GeV}
    \end{figure*}

\begin{figure*}
        \centering
        \includegraphics[width=0.9\linewidth]{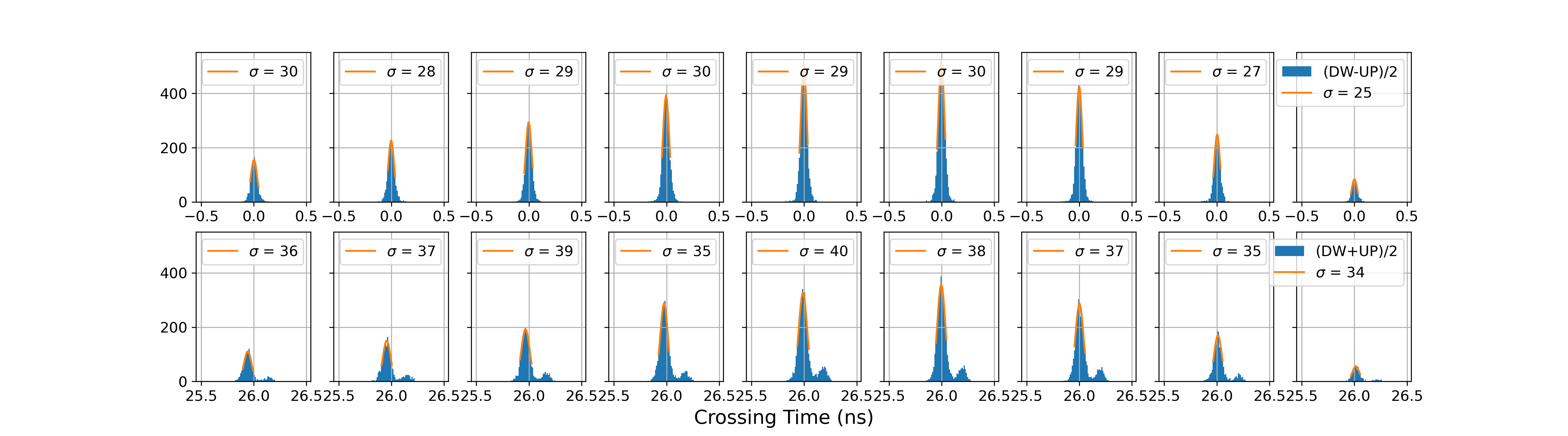}
        \includegraphics[width=\linewidth]{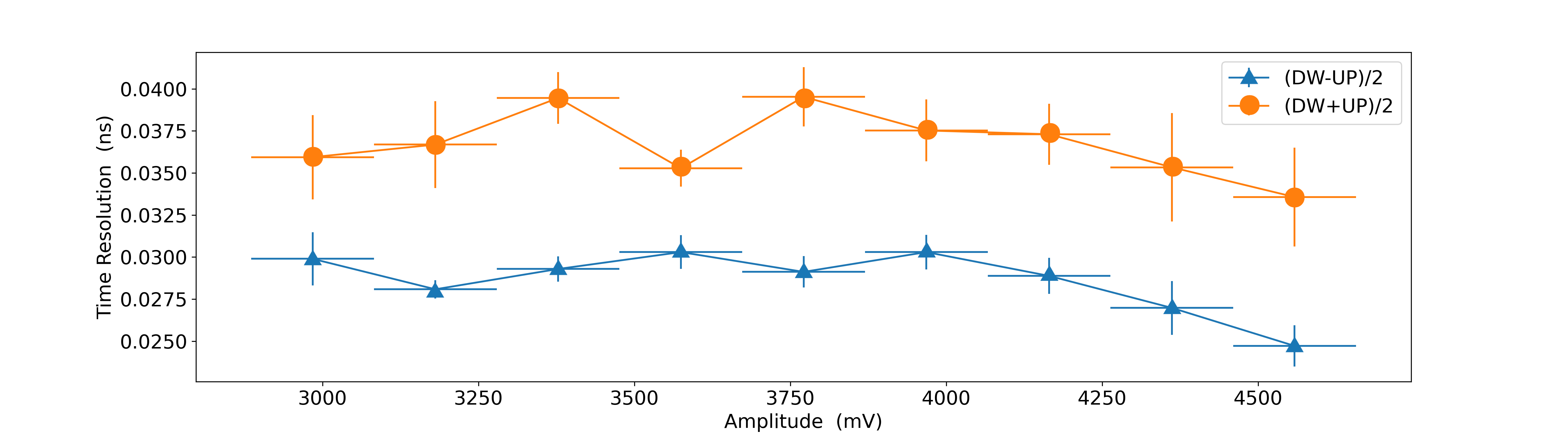}
        \caption{For E\textsubscript{beam} = 125 GeV. The measured energy E\textsubscript{meas} (in mV) has been subdivided into 9 bins, bin 1 at the left, bin 9 at the right in the lower figure. Top: the distributions of $(\Delta t_{DW} - \Delta t_{UP})/2$  for each bin of observed energy E\textsubscript{obs }in mV as indicated by the horizontal error bars in the lower plot. Middle: the distributions of $(\Delta t_{DW} + \Delta t_{UP})/2$ for each bin of observed energy. Lower: the measured time resolution $\sigma_t$ in ps for each bin of observed energy obtained from the standard deviation of the Gaussian fit to the corresponding timing peaks above in the figure. Blue values correspond to $(\Delta t_{DW} - \Delta t_{UP})/2$ and orange values correspond to $(\Delta t_{DW} - \Delta t_{UP})/2$.}
    \label{fig:125GeV}
    \end{figure*}

\begin{figure*}
        \centering
        \includegraphics[width=0.9\linewidth]{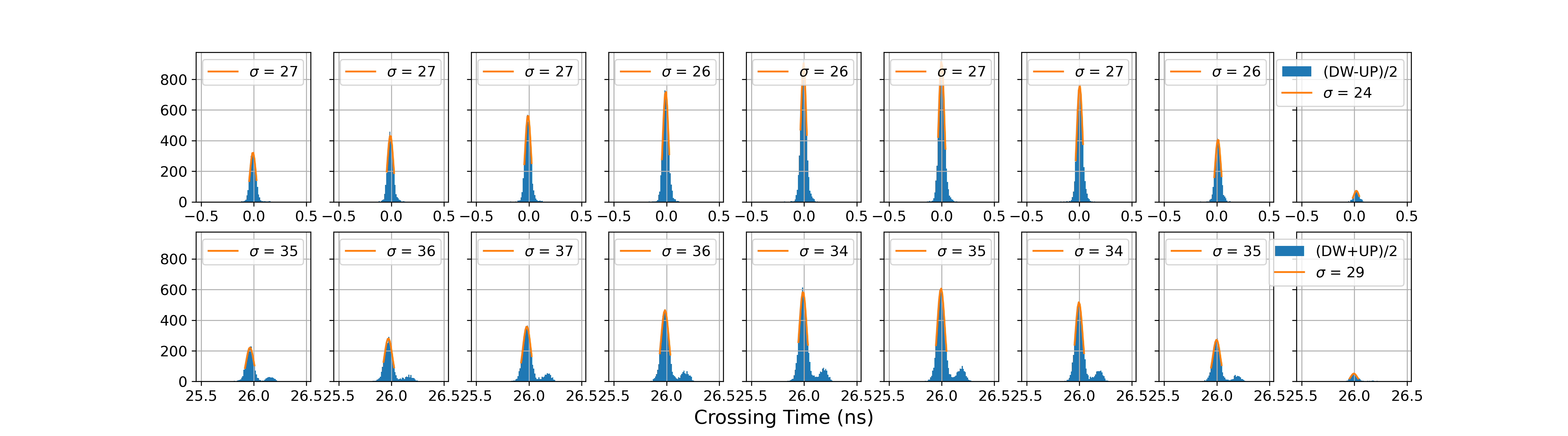}
        \includegraphics[width=\linewidth]{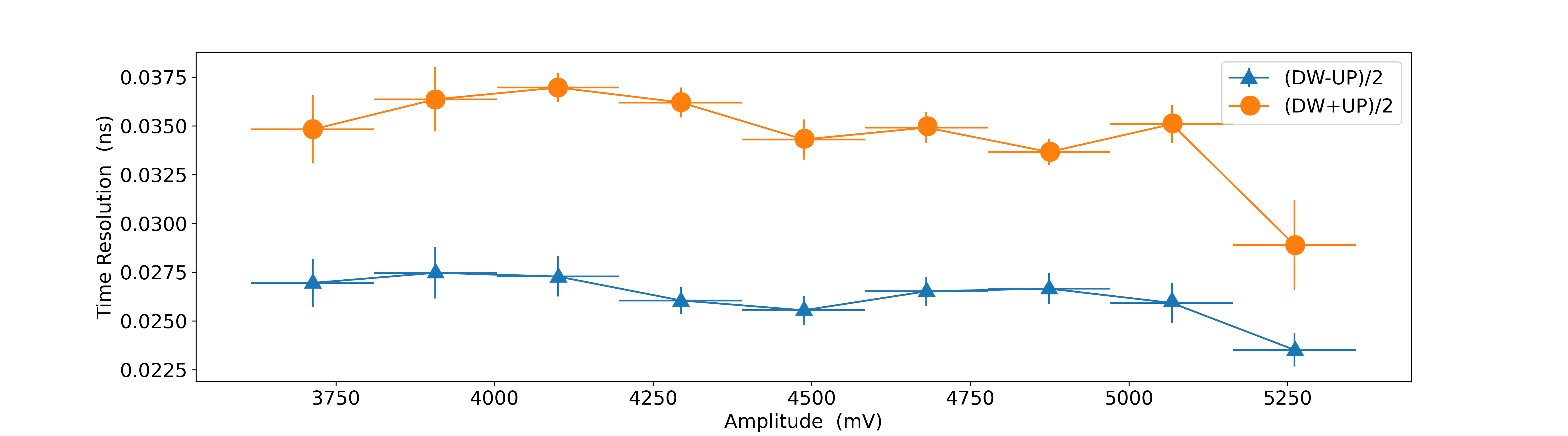}
        \caption{For E\textsubscript{beam} = 150 GeV. The measured energy E\textsubscript{meas} (in mV) has been subdivided into 9 bins, bin 1 at the left, bin 9 at the right in the lower figure. Top: the distributions of $(\Delta t_{DW} - \Delta t_{UP})/2$  for each bin of observed energy E\textsubscript{obs }in mV as indicated by the horizontal error bars in the lower plot. Middle: the distributions of $(\Delta t_{DW} + \Delta t_{UP})/2$ for each bin of observed energy. Lower: the measured time resolution $\sigma_t$ in ps for each bin of observed energy obtained from the standard deviation of the Gaussian fit to the corresponding timing peaks above in the figure. Blue values correspond to $(\Delta t_{DW} - \Delta t_{UP})/2$ and orange values correspond to $(\Delta t_{DW} - \Delta t_{UP})/2$.}
        \label{fig:150GeV}
    \end{figure*}

        For each step of electron beam energy $E_{beam}$, the corresponding measured energy distribution $E_{meas}$ in the RADiCAL module shown in Fig.~\ref{fig:nominalenergy} was divided into nine bins around the peak of the distribution, and the technique described in Section~\ref{E25Timing} to determine the timing resolution was applied bin-by-bin. These studies are shown in \Cref{fig:25GeV,fig:50GeV,fig:75GeV,fig:100GeV,fig:125GeV,fig:150GeV}. Displayed in each of the figures are the average of the timing difference between downstream and upstream averaged signals, $(\Delta t_{DW}~-~\Delta t_{UP})/2$, as well as the average of the sum of downstream and upstream averaged signals, $(\Delta t_{DW}~+~\Delta t_{UP})/2$. As noted in Section~\ref{MSTEM}, the former distribution is a direct measure of the RADiCAL timing resolution, whereas the latter distribution includes a non-negligible contribution due to the resolution of the MCP. The standard deviations of the Gaussian fits to these distributions were then plotted bin by bin as a function of measured energy for each of nine bins.

\section{Analysis summary and key results}
\label{DAC}

\begin{figure*}
        \centering
        \includegraphics[width=0.49\linewidth]{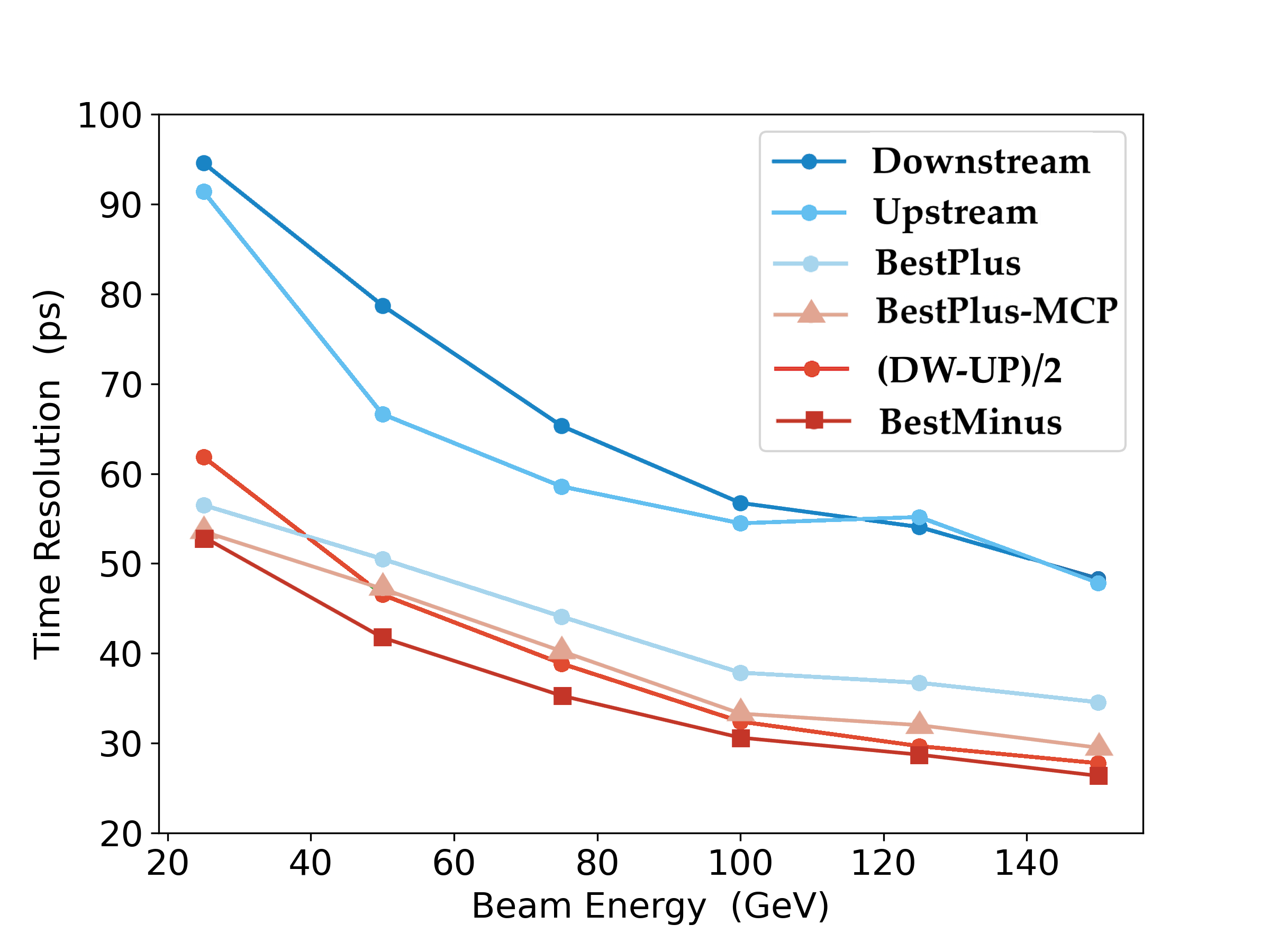}
        \includegraphics[width=0.49\linewidth]{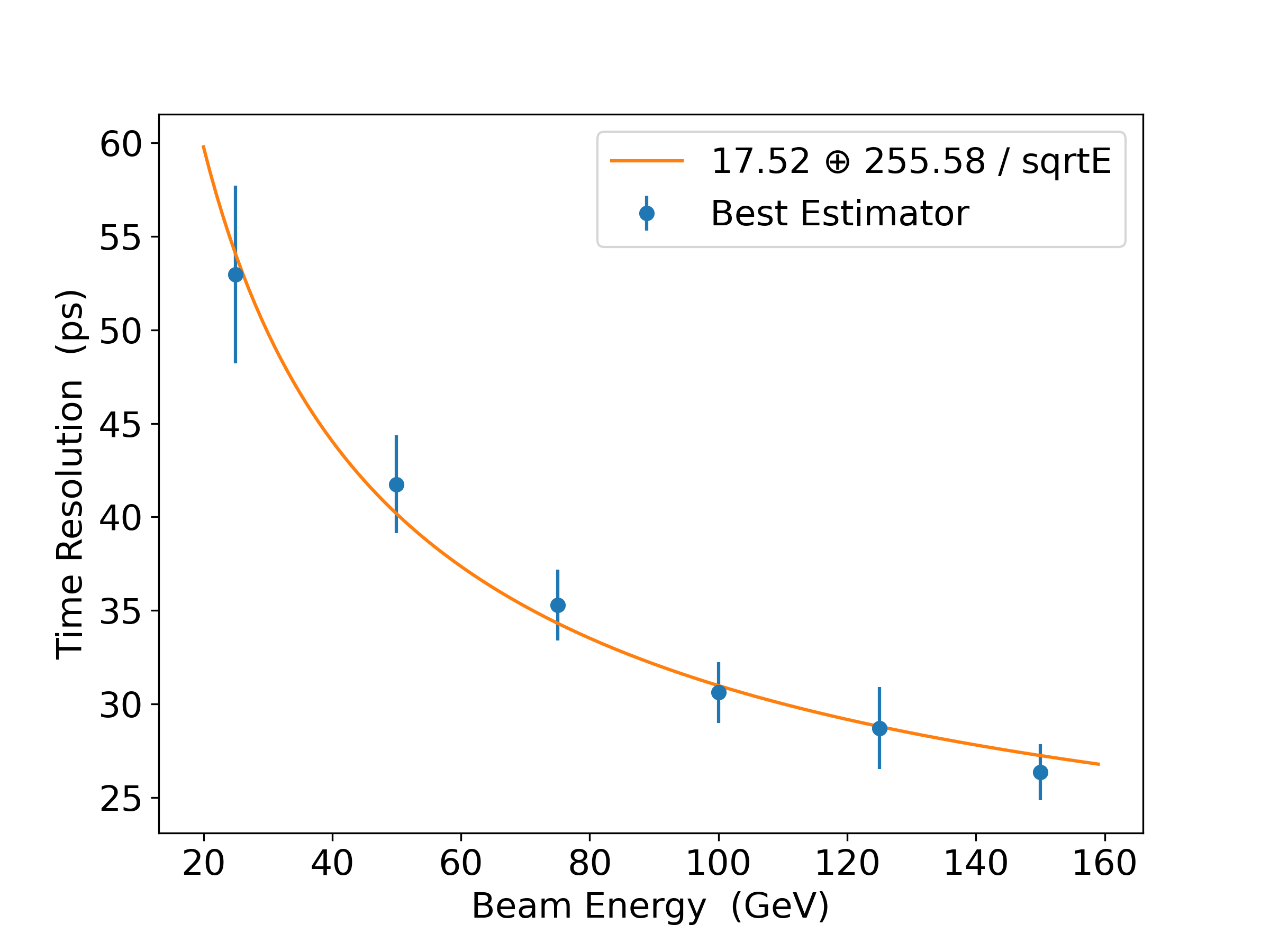}
        \caption{Left: Various methodologies for determining the timing resolution of the RADiCAL module (See Section~\ref{DAC}), with the legend identifying each method. Right: Best Estimator of the timing Resolution for the RADiCAL module, which is the BestMinus method, is displayed as a function of detected wnergy in the module for all six beam energies 25 GeV $\leq$ E $\leq$ 150 GeV.}
        \label{fig:finaltiming}
    \end{figure*}

    Given the extensive data sets recorded in the six energy steps of beam energy over the range 25 GeV $\leq$ E $\leq$ 150 GeV, the timing resolution of the RADiCAL module can be measured in a variety of ways. These methods are displayed graphically in Fig. 27 Left and are described here:

    \begin{enumerate}
	   \item \textit{Downstream in Fig.~\ref{fig:finaltiming} Left}. The module timing is determined from the average of the timing from each of four downstream SiPMs only, through the calculation of the quantity: $\Delta t_{DW} = \Sigma (t_{SiPM(i)}~-~t_{MCP})/4$ and fitting the peak of this distribution for which an example is shown in Fig.~\ref{fig:avgtiming} Upper Left for $E_{beam}$ = 25 GeV. For these calculations, the full spectrum of measured energy values corresponding to a given beam energy were used as shown in Figure~\ref{fig:nominalenergy}.
 
	   \item \textit{Upstream in Fig.~\ref{fig:finaltiming} Left}. The module timing is determined from the average of the timing from each of four upstream SiPMs only, through the calculation of the quantity: $\Delta t_{UP} = \Sigma (t_{SiPM(i)}~-~t_{MCP})/4$ and fitting the peak of this distribution for which an example is shown in Fig.~\ref{fig:avgtiming} Upper Right for $E_{beam}$ = 25 GeV. For these calculations, the full spectrum of measured energy values corresponding to a given beam energy were used as shown in Fig.~\ref{fig:nominalenergy}.
 
	   \item \textit{(DW-UP)/2 in Fig.~\ref{fig:finaltiming} Left}. The module timing is determined from the difference of the average timing of the downstream SiPMs relative to the MCP and the average timing of the upstream SiPMs relative to the MCP. The fits to these values are shown in Fig.~\ref{fig:timingdiffs}. Again, for these calculations, the full spectrum of of observed energy values corresponding to a given beam energy value were used as shown in Fig.~\ref{fig:nominalenergy}.
 
	   \item \textit{BestMinus in Fig.~\ref{fig:finaltiming} Left}. Here, as for Method 3 above, the module timing is determined from the difference of the average timing of the downstream SiPM relative to the MCP and the average timing of the upstream SiPM timing relative to the MCP, but using the detailed energy analysis shown in \Cref{fig:25GeV,fig:50GeV,fig:75GeV,fig:100GeV,fig:125GeV,fig:150GeV}. This analysis selects the measured time differences in bins 6-8 in each figure and computes the average of these time differences $(\Delta t_{DW}~-~\Delta t_{UP})/2$ to determine the timing resolution. This approach assures that the timing resolution is determined when the observed energy in the RADiCAL module corresponds very closely to the actual delivered beam energy.
 
	   \item \textit{BestPlus in Fig.~\ref{fig:finaltiming} Left}. Here the module timing is determined from the sum of the average of the downstream SiPM timing relative to the MCP and the upstream SiPM timing relative to the MCP, using the detailed energy analysis shown in \Cref{fig:25GeV,fig:50GeV,fig:75GeV,fig:100GeV,fig:125GeV,fig:150GeV}. This analysis selects the measured time averages in bins 6-8 in each figure and computes the overall average timing $(\Delta t_{DW}~+~\Delta t_{UP})/2$ to determine the timing resolution. This approach assures that the timing resolution is determined when the observed energy in the RADiCAL module corresponds closely to the actual delivered beam energy. However, this method includes an unsubtracted resolution contribution due to the MCP tube.
 
	   \item \textit{BestPlus-MCP in Fig.~\ref{fig:finaltiming} Left}. This analysis is identical to that of method 5 above, but subtracts a value of $\sigma_{t,MCP}~\approx$ 10 ps to correct for the presence of the MCP in the $(\Delta t_{DW}~+~\Delta t_{UP})/2$ approach to determination of the timing resolution of the module.
    \end{enumerate}

    Of these six approaches, method 4 (BestMinus in Fig.~\ref{fig:finaltiming} Left) provides the ``best estimator" of the timing resolution of the RADiCAL module in which the measured energy in the module $E_{meas}$ is effectively the same as the nominal beam energy $E_{beam}$ delivered to the module. Hence, we identify this methodology as the best estimator of the timing resolution of the RADiCAL module in its current form, and the results of method 4 are plotted in Fig.~\ref{fig:finaltiming} Right.

    The fit to the data points in Fig.~\ref{fig:finaltiming} Right indicates that the functional dependence of the timing resolution with respect to the energy measured in the RADiCAL module follows the form:

    \begin{equation}
	   \sigma_t =\frac{a}{\sqrt{E}}~\oplus~b
    \end{equation}
    where: a = $256\sqrt{GeV}$~ps,~b = 17.5~ps.
 
    Additionally, the time resolution measured at the highest electron beam energy for which data was currently recorded (150 GeV) was found to be $\sigma_{t} = 27$ ps. The data also indicates that in the limit of high energy, the timing resolution is $\sigma_{t} = 17.5$ ps under the current measurement conditions.

    \begin{figure}
        \centering
        \includegraphics[width=0.40\linewidth]{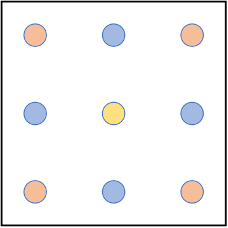}
        \hfil
        \includegraphics[width=0.40\linewidth]{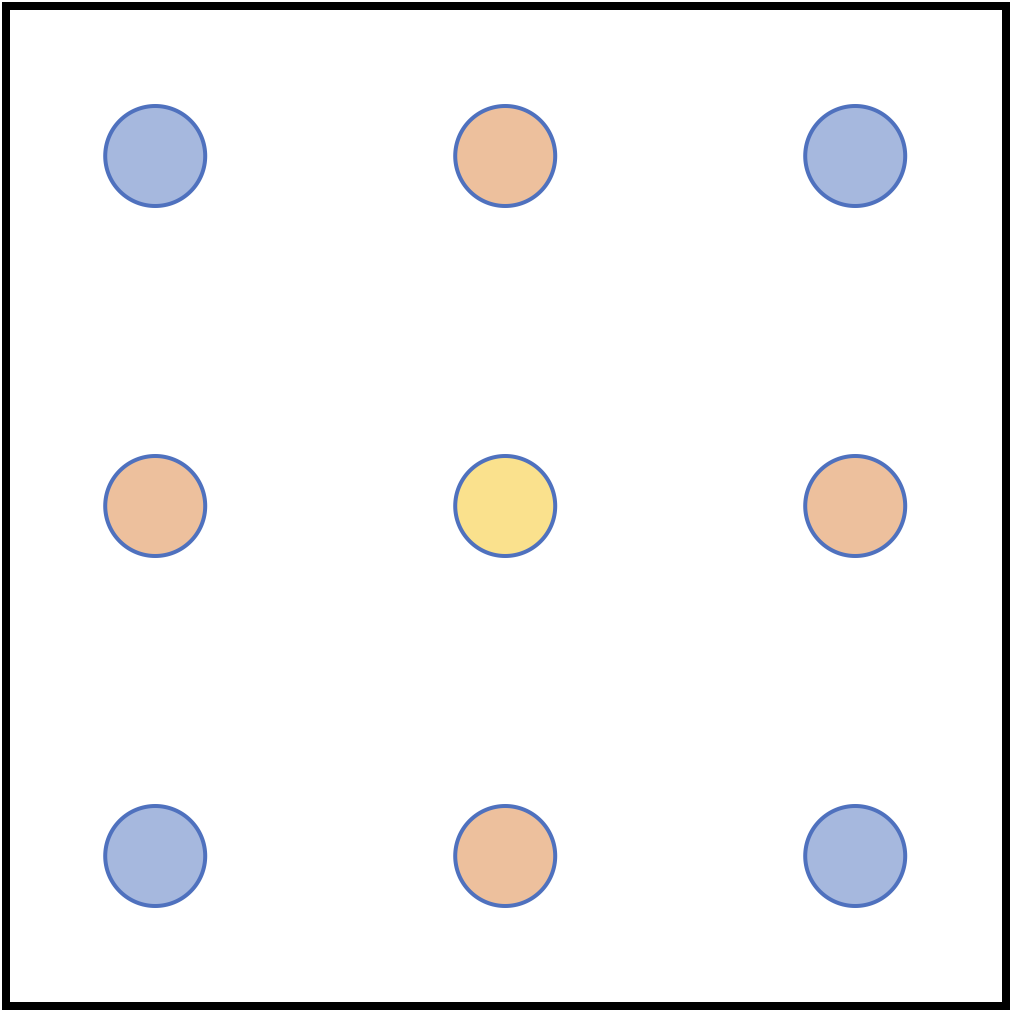}
        \caption{Potential capillary positioning in an enhanced RADiCAL module design that allows for simultaneous high-resolution timing and energy measurements. The lighter shaded  circles represent T-type capillaries, while the darker shaded circles represent E-type capillaries. The center circle houses a capillary that could either of T-type or E-Type, or alternatively could be used for calibration.}
    \end{figure}

\section{Ongoing research and development}
\label{section:ORD}

    The ongoing program of RADiCAL R\&D includes data analysis, instrumentation development, and simulation and beam testing activities.

    \begin {enumerate}

        \item \textit{Data Analysis: EM Shower Localization.} This is a systematic study of spatial localization of EM showers over the energy range 25 GeV $\leq$ E $\leq$ 150 GeV.  This analysis correlates energy information from the capillaries in the RADiCAL module with the spatial information of the incoming beam provided by a wire chamber positioned $\approx~{3~m}$ upstream of the RADiCAL module in the H2 beam line at CERN.

        \item \textit{Data Analysis: Comparison of DSB1 and LuAG:Ce Wavelength Shifters.} This is a comparative study of timing resolution of the RADiCAL module in which T-type capillaries containing filaments of the ceramic wavelength shifter LuAG:Ce are compared with the data presented here in which DSB1 organic fiber filaments were utilized.  This comparison will be in 25 GeV steps in electron beam energy over the range 25 GeV $\leq$ E $\leq$ 150 GeV.

        \item \textit{Instrumentation Development: Enhanced Modular Design.} Preparations are underway to improve the timing resolution and energy resolution of RADiCAL modules, through revised modular design, improved front-end electronics and Data Acquisition. Enhanced timing resolution can be realized by increasing the thickness of the LYSO:Ce crystal layers in the region of shower max by a factor of 2 to improve light collection.  Energy resolution can be enhanced by increasing the thickness of the LYSO:Ce crystal layers throughout the module to increase the sampling fraction for EM showers. This will reduce the stochastic term in the energy resolution to meet the FCC-hh requirements for EM energy resolution described in Sect. 1 above \cite{aleksa2019}.  The goals are to further improve (reduce) the timing resolution $\sigma_t$ of a single RADiCAL module by a factor of $\approx1/\sqrt{2}$ compared to its current value. To improve simultaneously the stochastic term in the overall energy resolution, by a similar factor, would require a  3 x 3 array of RADiCAL modules of updated design, as the full shower development is set by the Moli\'{e}re radius $R_{M}$ and a modular array is needed for transverse containment of the full EM shower \cite{ledovskoy2021-2}.  This is described next in this section.

        \item \textit{Simulation and Beam Test: Enhanced Modular Design.} A new modular structure with the potential to meet simultaneously the RADiCAL goals for both timing and energy resolution is currently under investigation through Geant4 simulation and is shown schematically in Fig. 28. In this module there are five T-type capillaries and four E-Type capillaries, each with SiPM readout at both ends and with high and low gain signals digitized. As the figure indicates, the capillaries can be deployed in various configurations to benefit timing, energy and positional analysis. For this structure, the E-Type capillaries will be filled with solid WLS filaments of LuAG:Ce ceramics running the full length of the module. LuAG:Ce is favored for its solid ceramic nature, ability to be drawn into filaments and excellent radiation hardness \cite{anderson2022,hu2020}. The T-Type capillaries will be filled with either DSB1 or LuAG:Ce WLS filaments for timing resolution comparison.  The cross section of the new module (18 x 18 mm$^2$) will be larger than that of the current module (14 x 14 mm$^2$) due to the increase in the radiation length and hence Moli\'{e}re radius of the new design.  This is a result of the thickness changes in the LYSO:Ce layers of the new structure. The intent is to conduct a comparative test of both the current and new modular structures for timing, energy and position localization in high energy electron beam over the energy range 25 GeV $\leq$ E $\leq$ 200 GeV. Once qualified in single module tests, a 3 x 3 array of such modules will be constructed and tested.
    \end{enumerate}

\section{Conclusions}
\label{conc}
    The results presented in Section~\ref{DAC}, and Fig.~\ref{fig:finaltiming} indicate that the RADiCAL modular calorimetry technique has the potential to meet and exceed the timing requirements for FCC-ee and FCC-hh applications, and additionally would be a potentially useful EM calorimetry approach for forward physics applications and fixed-target style experiments.

    Additionally, the impact of the RADiCAL research and development program, while directed toward future particle physics experimental applications, is not experiment specific, is potentially broad and significant, and can be expected to inform further developments in high energy physics instrumentation through ECFA DRD6 and CPAD9 programs. The technological innovations are versatile and can be applied in particle physics, nuclear physics, materials science, and  medical physics. 

\section{CRediT authorship contribution statement}
\label{CRedIT}
    \textbf{Carlos Perez Lara:} Conceptualization, Methodology, Software, Validation, Formal Analysis, Investigation, Supervision, Data Curation, Writing – Original Draft, Writing – Review and Editing. \textbf{James Wetzel:} Conceptualization, Methodology, Software, Validation, Formal Analysis, Investigation, Supervision, Data Curation, Writing – Original Draft, Writing – Review and Editing. \textbf{Ugur Akgun:} Resources. \textbf{Thomas Anderson:} Investigation, Resources. \textbf{Thomas Barbera:} Investigation, Formal Analysis. \textbf{Dylan Blend:} Investigation. \textbf{Kerem Cankocak:} Investigation, Supervision. \textbf{Salim Cerci:} Investigation, Writing – Review and Editing. \textbf{Nihal Chigurupati:} Investigation, Writing – Review and Editing. \textbf{Bradley Cox:} Conceptualization, Methodology, Validation, Investigation, Resources, Writing – Review and Editing, Supervision, Funding Acquisition. \textbf{Paul Debbins:} Investigation, Resources. \textbf{Max Dubnowski:} Investigation. \textbf{Buse Duran:} Investigation. \textbf{Gizem Gul Dincer:} Investigation. \textbf{Selbi Hatipoglu:} Investigation. \textbf{Ilknur Hos:} Investigation, Writing – Review and Editing. \textbf{Bora Isildak:} Investigation, Writing – Review and Editing. \textbf{Colin Jessop:} Funding Acquisition, Data Curation, Writing – Review and Editing. \textbf{Ohannes Kamer Koseyan:} Investigation. \textbf{Ayben Karasu Uysal:} Writing – Review and Editing. \textbf{Reyhan Kurt:} Writing – Review and Editing. \textbf{Berkan Kaynak:} Validation, Investigation, Resources, Writing – Review and Editing. \textbf{Alexander Ledovskoy:} Conceptualization, Methodology, Software, Validation, Formal Analysis, Investigation, Writing – Review and Editing. \textbf{Alexi Mestvirishvili:} Validation, Investigation, Supervision, Resources. \textbf{Yasar Onel:} Conceptualization, Validation, Investigation, Resources, Supervision, Funding Acquisition. \textbf{Suat Ozkorucuklu:} Investigation, Resources, Writing – Review and Editing, Supervision. \textbf{Aldo Penzo:} Investigation, Resources, Writing – Review and Editing. \textbf{Onur Potok:} Validation, Investigation, Resources, Writing – Review and Editing. \textbf{Daniel Ruggiero:} Validation, Investigation, Resources. \textbf{Randal Ruchti:} Conceptualization, Methodology, Validation, Investigation, Resources, Writing – Original Draft, Writing – Review and Editing, Visualization, Supervision, Project Administration, Funding Acquisition. \textbf{Deniz Sunar Cerci:} Investigation, Writing – Review and Editing. \textbf{Ali Tosun:} Investigation. \textbf{Mark Vigneault:} Investigation, Resources. \textbf{Yuyi Wan:} Investigation. \textbf{Mitchell Wayne:} Writing – Review and Editing, Project Administration, Funding Acquisition. \textbf{Taylan Yetkin:} Investigation, Writing – Review and Editing. \textbf{Liyuan Zhang:} Investigation, Resources, Writing – Review and Editing. \textbf{Renyuan Zhu:} Conceptualization, Methodology, Validation, Formal Analysis, Investigation, Resources, Writing – Review and Editing, Supervision, Funding Acquisition. \textbf{Caglar Zorbilmez:} Investigation, Writing – Review and Editing.

\section{Acknowledgements}
\label{acks}
    We thank the CERN SPS Coordination Team for their expert assistance in providing excellent beam conditions and logistics to make the experimental tests described above very successful. And We thank colleagues from the COMPASS experiment for the loan of Pb glass detectors which were used for backing calorimetry. 

    This is work has been supported in part by: US DOE grant DE-SC0017810, US NSF grant NSF-PHY-1914059, the University of Notre Dame through its Resilience and Recovery Grant Program, and QuarkNet for high school teacher and student participation.



  \bibliographystyle{elsarticle-num} 
  \bibliography{RADiCAL}




\end{document}